\documentclass[12pt,a4paper]{article}
\nonstopmode
\usepackage{epsfig}
\usepackage[a4paper]{geometry}
\usepackage{float}
\usepackage[stable]{footmisc}
\usepackage[utf8]{inputenc}
\usepackage{a4wide}
\usepackage{amsmath,amssymb,amstext,amsthm,amssymb}
\usepackage{amsfonts,mathrsfs}
\usepackage{framed}
\usepackage{graphicx}
\usepackage{color}
\usepackage[most]{tcolorbox}
\usepackage{bbold}
\usepackage{bbm}
\usepackage[normalem]{ulem}
\usepackage{rotating} %% rotating figures
\usepackage{slashed} %%feynman slash
\usepackage{cancel} %% diagonal line through word
\usepackage{braket} %% bra ket
\usepackage{amstext}
\usepackage{array}
\usepackage{setspace}
\usepackage[urlcolor=blue,colorlinks=true,linkcolor  = black]{hyperref}
\usepackage{overpic}
\usepackage{bbm}
\usepackage{cite}
\usepackage{ulem}
\usepackage{lipsum}
\usepackage{tikz}
\usepackage{mathtools}
\usepackage{amsthm}

\newcommand{\DeclareRuneSeparators}[1]{} %if needed copy the whole definition

\input{arm.fd}

\newcommand{\cK}{\mathcal{K}}

\usepackage{tikz-cd}
\usepackage{tikz}
\usepgflibrary{arrows} % LATEX and plain TEX and pure pgf 
\usetikzlibrary{arrows,shapes,positioning} % LATEX and plain TEX when using TikZ
\usetikzlibrary{decorations.markings}
\usetikzlibrary{shapes.geometric}
\usetikzlibrary{arrows.meta,arrows}

\tikzstyle arrowstyle=[scale=1]
\tikzstyle directed=[postaction={decorate,decoration={markings,
    mark=at position .5 with {\pgftransformscale{2}\arrow[arrowstyle]{stealth}}}}]

\let\baraccent=\=
\renewcommand{\=}[1]{\stackrel{#1}{=}}
\newcommand{\id}[0]{\mathbb{I}}
\newcommand{\cT}{\mathcal{T}}

\DeclareSymbolFontAlphabet{\mathbb}{AMSb}

\newcommand{\eq}[1]{\begin{align}#1\end{align}}

\renewcommand{\k}{\kappa}
\renewcommand{\L}{\Lambda}

\def\bbZ{\mathbb{Z}}
\def\bbQ{\mathbb{Q}}

\def\bbP{\mathbb{P}}
\def\bbF{\mathbb{F}}
\def\calK{\mathcal{K}}
\def\calC{\mathcal{C}}
\def\calI{\mathcal{I}}

\def\calH{\mathcal{H}}
\def\calV{\mathcal{V}}
\def\calS{\mathcal{S}}
\def\calC{\mathcal{C}}
\def\sat{{\text{sat}}}

\newcommand{\subeqslabel}[2]{\begin{subequations}\begin{align}#1\end{align}\label{#2}\end{subequations}}
\newcommand{\ignore}[1]{}

\begin{document}
	
	\pagestyle{plain}

	%----------------------------------------------------------------------%
	%  numbering equations with section number
	%----------------------------------------------------------------------%
	\makeatletter
	\@addtoreset{equation}{section}
	\makeatother
	\renewcommand{\theequation}{\thesection.\arabic{equation}}
	%----------------------------------------------------------------------%
	%  title page
	%----------------------------------------------------------------------%
	\pagestyle{empty}
	\begin{flushright}
        \text{CERN-TH-2023-189}
        \end{flushright}
	\vspace{0.5cm}
	 
	\begin{center}
		
		{\LARGE \bf{Counting Calabi-Yau Threefolds}
			\\[10mm]}
	\end{center}

	\begin{center}
		\scalebox{0.95}[0.95]{{\fontsize{14}{30}\selectfont Naomi Gendler,$^{a,b}$ Nate MacFadden,$^{b}$ Liam McAllister,$^{b}$ Jakob Moritz,$^{b,c}$}} \vspace{0.35cm}
		\scalebox{0.95}[0.95]{{\fontsize{14}{30}\selectfont Richard Nally,$^{b}$ Andreas Schachner,$^{b,d}$ and Mike Stillman$^{e}$}}
	\end{center}

	\begin{center}
		\vspace{0.25 cm}
		\textsl{$^{a}$Jefferson Physical Laboratory, Harvard University, Cambridge, MA 02138 USA}\\
		\textsl{$^{b}$Department of Physics, Cornell University, Ithaca, NY 14853 USA}\\
        \textsl{$^c$Department of Theoretical Physics, CERN, 1211 Meyrin, Switzerland}\\
        \textsl{$^{d}$ASC for Theoretical Physics, LMU Munich, 80333 Munich, Germany}\\
		\textsl{$^{e}$Department of Mathematics, Cornell University, Ithaca, NY 14853 USA}\\

		\vspace{1cm}
		\normalsize{\bf Abstract} \\[8mm]

	\end{center}
    We enumerate topologically-inequivalent compact Calabi-Yau threefold
    hypersurfaces. By computing arithmetic and algebraic invariants and the Gopakumar-Vafa invariants of curves, we prove that the number of distinct simply connected Calabi-Yau threefold hypersurfaces resulting from triangulations of four-dimensional reflexive polytopes is 4, 27, 183, 1{,}184 and 8{,}036 at $h^{1,1} =1, 2, 3, 4,\ \text{and} \ 5$, respectively. We also establish that there are ten equivalence classes of Wall data of non-simply connected Calabi-Yau threefolds from the Kreuzer-Skarke list.  Finally, we give a provisional count of threefolds obtained by enumerating non-toric flops at $h^{1,1} =2$.
    
	\begin{center}
		\begin{minipage}[h]{15.0cm}

		\end{minipage}
	\end{center}
	\newpage
	%----------------------------------------------------------------------%
	%  Resetting of counters
	%----------------------------------------------------------------------%
	\setcounter{page}{1}

    \setcounter{tocdepth}{1}

	\pagestyle{plain}
	\renewcommand{\thefootnote}{\arabic{footnote}}
	\setcounter{footnote}{0}
	%----------------------------------------------------------------------%
	%  Paper begins
	%----------------------------------------------------------------------%
	%
	%
	\tableofcontents
	\newpage

\section{Introduction}
\label{sec:intro}

The landscape of four-dimensional effective field theories arising from Calabi-Yau compactifications of critical string theory is immensely rich, in part because of
the vast number of possible compactification geometries. 
To understand the size of this landscape, one needs to determine the number of Calabi-Yau threefolds.
More precisely, one should count threefolds\footnote{Throughout this paper, \emph{threefold} (or synonymously, \emph{phase}) will always mean \emph{compact Calabi-Yau threefold}, unless specified otherwise.} that give rise to inequivalent effective theories; this entails counting topological equivalence classes of threefolds.

Despite the importance of this problem, there presently exists no effective strategy for solving it. 
Wall's theorem \cite{wall} has provided a framework,
by establishing that simply connected Calabi-Yau threefolds with torsion-free homology are completely classified
by what we will call \emph{Wall data}: the
Hodge numbers $h^{1,1}$ and $h^{2,1}$, the triple intersection numbers $\k_{ijk}$, and the second Chern class $c_2$.  Two simply connected threefolds with equivalent Wall data are homeomorphic, and in fact even diffeomorphic, and give rise to equivalent effective theories upon compactification.
However, applying Wall's theorem in practice remains a
difficult computational challenge:
given the Wall data for a pair of 
potentially equivalent threefolds, checking equivalence requires either finding a basis transformation $\Lambda \in \text{GL}(h^{1,1},\mathbb{Z})$ that maps the Wall data of the pair into each other, or else proving that no such transformation exists.
Without any further information, the cost of such a search is exponential in $h^{1,1}$, and so the task rapidly becomes infeasible.

The purpose of this work is to identify and count equivalence classes of threefolds.
Our strategy is to compute \emph{invariants of the Wall data} --- a collection of arithmetic and algebraic invariants enumerated in \S\ref{sec:topology} and summarized in Table~\ref{tab:wallinvts} --- and to use these, as well as the Gopakumar-Vafa (GV) invariants \cite{Gopakumar:1998ii,Gopakumar:1998jq} of curves, to simplify the process of classification.

We will consider two large categories of threefolds. The first consists of 
hypersurfaces in toric fourfolds, which we refer to as \emph{toric phases}.
Any fine, regular, star triangulation (FRST) of a four-dimensional reflexive polytope defines a toric variety in which the generic anticanonical hypersurface is a smooth Calabi-Yau \cite{Batyrev:1993oya}.
The Kreuzer-Skarke database \cite{Kreuzer:2000xy} is a complete list of all reflexive polytopes in four dimensions.  Each such polytope admits a number of FRSTs that grows exponentially with the number of points in the polytope, and is as large as $10^{928}$ in an extreme example \cite{Demirtas:2020dbm}.  However, it is known that the number of genuinely inequivalent threefolds that can be constructed in this way is many orders of magnitude smaller than the total number of FRSTs \cite{Demirtas:2020dbm}. We aim to determine how much smaller this number is.

A main result of this work is an exact count of the number of equivalence classes of Wall data in the case of toric phases with $1 \le h^{1,1} \le 5$: see Table \ref{tab:toricSummary}.
We additionally bound the number of equivalence classes of Wall data of toric phases at $h^{1,1}=6\text{ and }7$: see Table \ref{tab:BoundsToricFav67}.

To turn a count of equivalence classes of Wall data to a count of topological equivalence classes of threefolds, one needs to be able to apply Wall's theorem, which  assumes that the threefold is simply connected. We find that there are ten equivalence classes of Wall data of non-simply connected threefolds arising from the Kreuzer-Skarke list. Setting these ten  classes aside and considering only the simply connected threefolds that remain, we apply Wall's theorem to prove that the number of topological equivalence classes of simply connected toric phases is 4 at $h^{1,1}=1$, 27 at $h^{1,1}=2$, 183 at $h^{1,1}=3$, 1184 at $h^{1,1}=4$, and 8036 at $h^{1,1}=5$.  
These counts are definite, and should be thought of as a theorem: no assumptions are made in their calculation.

\begin{table}
\centering
\resizebox{\columnwidth}{!}{
\begin{tabular}{|c|c|c|c|c|c|}
\hline 
 &  &  &  &  &    \\ [-0.9em]
$h^{1,1}$ & {\#} polys & {\#} FRSTs & {\#} FRST classes  & {\#}  CYs with $\pi_1=0$ & {\#} ECs with $\pi_1 \neq 0$ \\ [0.2em]
\hline 
\hline
 &  &  &  &  &   \\ [-0.9em]
1 & 5+0 & 5+0 & 5+0 & 4+0 & 1+0  \\ [0.2em]
\hline 
 &  &  &  &  &    \\ [-0.9em]
2 & 36+0 & 48+0 & 36+0 & 27+0 & 2+0   \\ [0.2em]
\hline 
 &  &  &  &  &    \\ [-0.9em]
3 & 243+1 & 525+1 & 274+1 & 183+0 & 3+0    \\ [0.2em]
\hline 
 &  &  &  &  &    \\ [-0.9em]
4 & 1,185+12  & 5,330+18  & 1,760+14 & 1183+1 & 3+0  \\ [0.2em]
\hline 
 &  &  &  &  &   \\ [-0.9em]
5 & 4,897+93 & 56,714+336 & 11,713+134 & 8,016+20 & 0+1   \\ [0.2em]
\hline 
\end{tabular} 
}
\caption{Main results for counts of topological equivalence classes of Calabi-Yau threefold
hypersurfaces in toric varieties, denoted `CYs' in this table.
Numbers of the form $M+N$ denote contributions of favorable and non-favorable polytopes, respectively. An FRST class or CY class is defined as non-favorable if it has no favorable representative. Counts labeled `ECs' are of equivalence classes of Wall data with $\pi_1 \neq 0$.}\label{tab:toricSummary}
\end{table}

Our results for toric phases are summarized in Table \ref{tab:toricSummary}.  For each value of $h^{1,1}$, we list the number of polytopes, 
the number of FRSTs, and the number of FRST classes, defined as sets of FRSTs that agree up to polytope automorphism when restricted to two-faces.
We then give the numbers of equivalence classes of Calabi-Yau threefold hypersurfaces in toric varieties.
The counts are given as $M+N$, with $M$ the number of favorable classes and $N$ the number of non-favorable classes; we call a class favorable if at least one representative of the class arises from a triangulation of a favorable polytope.\footnote{A polytope $\Delta^{\circ}$ is called favorable if every two-face of $\Delta^{\circ}$ that has interior points is dual to a one-face of the dual polytope $\Delta$ having no interior points. Computing the topological data of a threefold resulting from a triangulation of a polytope is comparatively simple when $\Delta^{\circ}$ is favorable.}
 
The second category of threefolds that we consider are those that can be constructed by performing flops from a toric threefold. 
As we will show, some such flops result in threefolds that 
are not topologically equivalent to a hypersurface in a toric variety.
We call such threefolds \textit{non-toric phases}.

We construct non-toric phases following the algorithm presented in \cite{Gendler:2022ztv}, which makes use of the topological data and GV invariants of some chosen toric phase.
Because we compute GV invariants only up to a finite cutoff, 
it is not certain that we will succeed in identifying all non-toric phases. We therefore obtain provisional counts of the number of non-toric phases; in particular, we find six non-toric phases at $h^{1,1}=2$. In contrast to our counts of toric phases, this number does not represent a theorem; we will review the assumptions underlying its computation in \S\ref{sec:non-toric}.

Having constructed a provisionally complete set of threefolds, both toric and non-toric, we turn to our goal of partitioning this set into topological equivalence classes.  
We begin by splitting the set of threefolds at fixed Hodge numbers into subsets of potentially equivalent phases, for which certain invariants computed from the Wall data agree.  The invariants we use are enumerated in Table \ref{tab:wallinvts}. We then use GV invariants to guess the form of basis transformation matrices between members of each set of potentially equivalent phases.\footnote{In the related context of complete intersection Calabi-Yau threefolds (CICYs), progress has been made in e.g.~\cite{He:1990pg,Anderson:2008uw,Carta:2021sms} by brute-force computation. These analyses proceeded by searching in a box in $\mathbb{Z}^{(h^{1,1})^2}$ for the  basis transformations demanded by Wall's theorem, and, while they were able to find some equivalences, they were unable to compute the exact number of phases.} In this way, we are able to compute upper and lower bounds on the number of phases and, when these bounds coincide, we find the exact number of phases.

This paper is organized as follows. In \S\ref{sec:topology}, we review Wall's theorem and define the topological invariants used in our analysis. In \S\ref{sec:toric} and \S\ref{sec:non-toric} we present our results for toric and non-toric phases, respectively. We conclude in \S\ref{sec:conclusions}.

\textbf{Note added:} After this paper was completed we received \cite{oxford}, which likewise counts diffeomorphism classes of Calabi-Yau threefolds. Although our approaches are similar in spirit, we employ different invariants; we include both favorable and non-favorable polytopes; we study threefolds obtained via non-toric flops; and we sub-classify according to $\pi_1$.
Moreover, although \cite{oxford} claims to restrict to
simply connected threefolds, not all cases 
considered in \cite{oxford} are simply connected (cf.~\cite{Batyrev:2005jc}).
Finally, some of the results of \cite{oxford} are compatible with our findings, while others are not.
Exact results are obtained in \cite{oxford} for three cases: for favorable polytopes at $h^{1,1}=1$, $2$, and $3$ they find 5, 29, and 186 equivalence classes, and --- provided we combine our counts with $\pi_1=0$ and $\pi_1 \neq 0$, see Table \ref{tab:toricSummary} --- we confirm these counts.  For favorable polytopes at $h^{1,1}=4$,  \cite{oxford} quotes an upper bound of 1185 equivalence classes, but we find an exact count of 1186.
For favorable polytopes at $h^{1,1}=5$, \cite{oxford} finds a range, and we find an exact count compatible with this range, while at 
$h^{1,1}=6$, \cite{oxford} finds a lower bound 54,939, which is incompatible with our upper bound 54,141.  It would be worthwhile to resolve these discrepancies.\footnote{There is also a minor discrepancy in the counts of favorable polytopes: in \cite{oxford} the number of such polytopes is quoted as $4{,}896$ and $16{,}607$  at $h^{1,1}=5$ and $6$, whereas we find $4{,}897$ and $16{,}608$, but our counts of FRSTs agree exactly with those of \cite{oxford}.}

\section{Topological invariants of Calabi-Yau threefolds}
\label{sec:topology}

The main goal of this work is to classify topologically equivalent threefolds at small $h^{1,1}$.  
The foundational tool is Wall's theorem \cite{wall}: 
\newtheorem*{theorem*}{Theorem (Wall)}
\begin{theorem*}\label{eq:wallstheorem}
The homotopy type of a compact, simply connected Calabi-Yau threefold with torsion-free homology is completely determined by its Hodge numbers, triple intersection numbers, and second Chern class.
\end{theorem*}
To understand how to apply this theorem, we let $X$ be a 
Calabi-Yau threefold, 
with $\kappa$ its triple intersection form
\begin{equation}
    \kappa:\, H^2(X,\mathbb{Z})\times H^2(X,\mathbb{Z})\times H^2(X,\mathbb{Z})\rightarrow \mathbb{Z}\, ,
\end{equation}
and $c_2(TX)\in H_2(X,\mathbb{Z})$ its second Chern class, naturally viewed as a map
\begin{equation}
    c_2:\, H^2(X,\mathbb{Z})\rightarrow \mathbb{Z}\, .
\end{equation}
Then, in a lattice basis $\{D_i\}$ of $H^2(X,\mathbb{Z})\simeq \mathbb{Z}^{h^{1,1}}$, we may express the data of $\kappa$ and $c_2$ as a totally symmetric three-tensor $\kappa_{ijk}$, and a linear form $c_{2,i}$ defined as
\begin{equation}
    \kappa_{ijk} = \int_{X}\, D_i\wedge D_j\wedge D_k\; ,\quad c_{2,i} = \int_{D_i}\, c_2(X) .
\end{equation} 
We refer to $h^{1,1}$, $h^{2,1}$, $\kappa$, and $c_2$ collectively as the \emph{Wall data} of $X$.

Now suppose that $X, X'$ are two compact threefolds with the same Hodge numbers.  At this stage we do not require that $X$ and $X'$ are simply connected and have torsion-free homology.
We say that $X$ and $X'$ are \emph{Wall-equivalent}
if there exists an integral change of basis 
$\L\in\operatorname{GL}(h^{1,1},\bbZ)$ such that 
\begin{equation}\label{eq:LambdaDef}
    {\kappa^{\prime}}_{i'j'k'} = {\L^i}_{i'}{\L^j}_{j'}{\L^k}_{k'} \k_{ijk}, \ \ \  {c^\prime}_{2,i'} = {\L^i}_{i'} c_{2,i}\, ,
\end{equation}
where $(\kappa_{ijk},c_{2,i})$ and $({\kappa^\prime}_{ijk}, {c^\prime}_{2,i})$ denote the triple intersection forms and second Chern classes of $X$ and ${X^\prime}$, respectively, given in some arbitrary integral bases.  Importantly, we require the basis transformation matrix $\L$ to be invertible over the integers, so its determinant must be $\pm1$. 

Wall's theorem implies that if two compact, simply connected, torsion-free threefolds $X$, $X^\prime$ are Wall-equivalent, then they are topologically equivalent; specifically, they are both homeomorphic and diffeomorpic. 
One of the main results of this work is an exact count of the equivalence classes of Wall data for all Calabi-Yau threefold hypersurfaces with $h^{1,1} \le 5$ obtained from the Kreuzer-Skarke list, shown in Table \ref{tab:toricSummary}.

\subsection{Invariants of Wall data}
\label{sec:wall_invts}

Much of the progress to date in proving equivalences between threefolds has centered around brute force searches for a basis transformation $\L$ as in \eqref{eq:LambdaDef}, e.g.,~by searching over matrices $\L \in \operatorname{GL}(h^{1,1},\bbZ)$ whose entries are bounded above by some parameter as in \cite{He:1990pg,Carta:2021sms}. This approach rapidly becomes expensive, and
we will proceed in a different manner.

An easy way to determine whether a threefold with Wall data $(\k,c_2)$ is Wall-\emph{inequivalent} to a threefold with Wall data $(\k',c_2')$ is to compute invariants of the Wall data; if any such invariants differ, then the threefolds are inequivalent. In this section we will introduce the invariants used in our analysis, which will be defined in terms of $\k$ and $c_2$, as well as the rank-four tensor $H_{ijkl}$ defined in \cite{Hubsch:1992nu} as \eq{H_{ijkl} = -2\left(\k_{ijk}c_{2,l} + \text{cyclic permutations}
\right).}

\subsubsection{Divisibility invariants}

Perhaps the simplest invariants are \textit{divisibility invariants}, defined in terms of the numerical entries of $\k$ and $c_2$. The gcd of an integer vector is invariant under the action of $\operatorname{GL}(n,\bbZ)$, and so \eq{d_0 \coloneqq \gcd(c_2)\label{eq:divinvd0}} is invariant under the change of basis used in Eq.~\eqref{eq:LambdaDef}. 

Similarly, \cite{Hubsch:1992nu} defines a series of seven divisibility invariants from the entries of the higher-rank tensors, three from $\k$ and four from $H$. In any basis for $H_4(X)$, these are defined as\footnote{The formula for $d_7$ is misprinted in \cite{Hubsch:1992nu}; we have corrected it here.}
\subeqslabel{d_1 &\coloneqq \gcd\left(\k_{ijk}\right)\label{eq:divinvhubsch1} \\ 
d_2 &\coloneqq \gcd\left(\k_{iij},2\k_{ijk}\right) \\
d_3 &\coloneqq \gcd(\left[\k_{iii},3\left(\k_{iij}\pm\k_{ijj}\right),6\k_{ijk}\right] \\
d_4 &\coloneqq \gcd\left(H_{ijkl}\right)\\
d_5 &\coloneqq \gcd\left(H_{iijk}, 2H_{ijkl}\right)\\
d_6 &\coloneqq \gcd\left[H_{iiij}, 3\left(H_{iijk} \pm H_{ijjk}\right),6H_{ijkl}\right]\\
d_7 &\coloneqq \gcd\left[H_{iiii},2\left(2H_{iiij} \pm 3H_{iijj} \pm 2 H_{ijjj}\right), 12\left(H_{iijk} \pm H_{ijjk} \pm H_{ijkk}\right), 24H_{ijkl}\right],}{eq:hubsch} where indices are \textit{not} summed over.
 
These invariants obey a hierarchical pattern of divisibility: $d_1$ divides $d_2$, $d_2$ divides $d_3$, and so on; however, $d_0$ need not divide the others. We thus have a vector $d$ of eight integers that must be the same in any two threefolds that are topologically equivalent.

\subsubsection{Arithmetic invariants}

The zero locus of any finite set of homogeneous polynomials defines a (possibly singular) projective variety; if the coefficients of the polynomials are rational numbers, then these will be varieties defined over $\bbQ$, which in slightly nonstandard terminology we refer to as \textit{rational varieties}. Accordingly, because the entries of $\k_{ijk}$ and $c_2$ are integers, we can define projective hypersurfaces in $\bbQ\bbP^{h^{1,1}-1}$ in terms of  a vector $x^i$ of formal variables\footnote{We will not distinguish between raised and lowered indices for these formal variables.} as\footnote{Throughout this section, calligraphic font will be used to denote varieties.} 
\begin{enumerate}
\item The cubic surface $\calK$ defined by the equation \eq{\k_{ijk}\, x^ix^jx^k=0 \subset \bbQ\bbP^{h^{1,1}-1}\,.}
\item The codimension-two surface $\calI$ defined by the intersection of
$\calK$
with the hyperplane $c_{2,i}\, x^i=0\subset \bbQ\bbP^{h^{1,1}-1}$
, i.e., the set of points simultaneously satisfying \eq{c_{2,i}\, x^i = \k_{ijk}\, x^ix^jx^k=0 \subset \bbQ\bbP^{h^{1,1}-1}\,.}
\item The quartic surface $\calH$ defined by the equation \eq{H_{ijkl}\, x^ix^jx^kx^l=0 \subset \bbQ\bbP^{h^{1,1}-1}\,.}
\end{enumerate}
 
A basic invariant constructed from these varieties is their smoothness. A variety $\mathcal{V}$ defined by the vanishing of polynomials $f_i$ is said to be singular if there exists a point on $\calV$ such that the Jacobian matrix $J = \partial_j f_i$ has rank smaller than the codimension of $\mathcal{V}$, and is called smooth otherwise. For any particular threefold, the varieties $\calK$ and $\calI$ may or may not be smooth; however, whenever $\calI$ is non-empty, and so in particular for all threefolds with $h^{1,1}>2$, the variety $\calH$ is necessarily singular. This can be seen from the fact that the defining polynomial $H_{ijkl}x^ix^jx^kx^l$ factorizes \eq{H_{ijkl}x^ix^jx^kx^l = -8\left(\k_{ijk}x^ix^jx^k\right)\left(c_{2,l}x^l\right).}

The varieties $\mathcal{I}$, $\mathcal{K}$, and $\mathcal{H}$ are rational varieties, i.e. their defining polynomials have integer coefficients. Moreover, the change of basis in Eq.~\eqref{eq:LambdaDef} defines an isomorphism at the level of rational varieties.\footnote{The converse, however, is not necessarily true. Two threefolds with isomorphic associated varieties $\calK,\ \calI$, and $\calH$ need not in general be Wall-equivalent, because change of basis matrices in $\operatorname{GL}(h^{1,1},\bbQ)$ are sufficient for isomorphism in the category of rational varieties but not for Wall equivalence. We will shortly encounter an example where this distinction is important.} Thus, \emph{the varieties 
$\calK,\ \calI$, and $\calH$ 
are themselves Wall invariants.} It is fruitful to study these varieties through the lens of arithmetic geometry, which
allows us to define further invariants.\footnote{For more on arithmetic geometry, and in particular point counts, in the context of string compactifications, see e.g.~\cite{Candelas:2019llw,Kachru:2020sio,Kachru:2020abh,Candelas:2023yrg}.}

Let us begin by briefly reviewing the basic context of arithmetic geometry; more details can be found in e.g. \cite{Ribet,yui:review,meyer:book}. Consider a rational variety $\calV \subset \bbP^{n-1}$. From this variety, we can can build an infinite number of related varieties, one defined over each finite field. 

Recall that, for each prime number $p$, $\bbZ/p\bbZ$ defines a finite field $\bbF_p$, where addition and multiplication are taken in $\bbZ$ and then reduced modulo $p$. For instance, in $\bbF_3$ we have $2+1=0$ and $2\times2=1$ since 3 and 4 are congruent to 0 and 1 modulo 3, respectively. By reducing the coefficients of the defining polynomial of $\calV$ modulo $p$ and then restricting its variables to only take values in $\bbF_p$, we can build a variety $\calV_p$ over $\bbF_p$. For instance, the rational variety $x_1^3+3x_2^3 +4 x_3^3- 5x_1x_2x_3=0$ becomes, when reduced over $\bbF_3$, $x_1^3+x_3^3+x_1x_2x_3=0$. 

We are now in a position to define our first arithmetic invariants, namely \emph{bad primes}. Let us take $\calV$ to be smooth. When reduced modulo a prime $p$, the resulting variety $\calV_p$ can be singular even though $\calV$ is smooth. In general, this can occur for only finitely many primes, called the bad primes of $\calV$. As an example, consider the smooth rational variety $x_1^3+x_2^3+x_3^3=0$; when reduced over $\bbF_3$, this is a singular variety, as can be readily seen by evaluating its derivatives at e.g.~the point $x^i = (1,2,0)$, and so 3 is a bad prime for this variety. The bad primes of $\calK$ and $\calI$ are thus invariants of the Wall data.\footnote{Recall that $\calH$ is singular for $h^{1,1}>2$, and thus does not have bad primes as such.} We describe a systematic way to compute the bad primes of any rational variety in the next subsection. 

Each of the prime-order fields $\bbF_p$ defined above admits an infinite number of field extensions $\bbF_{p^s}$ defined by adjoining the root of an irreducible order-$s$ polynomial. Thus, from $\bbF_2$ we can build $\bbF_4$, $\bbF_8$, and so on. In fact, these are the \textit{only} finite fields. Moreover, $\bbF_p$ embeds directly into each of these field extensions. Therefore, by simply declaring our variables $x_i$ to now live in $\bbF_{p^s}$ rather than just $\bbF_p$, we can build from each of the infinite number of varieties over $\bbF_p$ constructed above an additional infinite tower of varieties, $\calV_{p^s}$. 

Now let $\calV_{p^s}$ be such a variety. There are only a finite number of points on $\calV_{p^s}$ because the variables appearing in the defining polynomial take values in $\bbF_{p^s}$. 
We define the point counts
\begin{equation}\label{eq:ptcounts}
    \#\bigl(\calV,p^s\bigr) := \#\Bigl( x \in \bbF_{p^s} \mathbb{P}^{n-1} \Bigl|\, x \in \mathcal{V}\Bigr)
\end{equation}
The $\#(\calV,p^s)$
are extremely powerful invariants, and are a basic object of study in arithmetic geometry.
Applied to our set of varieties, we compute point counts on $\calK$ and $\calI$. 
We note that, although point counts are efficiently computable at small $h^{1,1}$ and over finite fields of small order, the computation becomes slow at large $h^{1,1}$ and at large $p^s$.

As an aside, we note some interesting simplifications that occur at $h^{1,1}=3$. In this case, $\k_{ijk}x^ix^jx^k$ is a cubic in three variables, and so, if $\calK$ is smooth,
then by the adjunction formula $\calK$ is a Calabi-Yau one-fold, i.e.,~an elliptic curve. It is a classical result that any cubic elliptic curve admits an essentially unique representative in Weierstrass form, i.e., as a choice of rational coefficients $f,\ g$ in the defining polynomial \eq{y^2=x^3+fx+g.} Two cubics are isomorphic as rational varieties if and only if they have the same Weierstrass representative, so rather than computing e.g.~point counts and bad primes, at $h^{1,1}=3$ 
one could 
classify the potentially equivalent choices of $\k_{ijk}$ at fixed $h^{2,1}$ by finding the Weierstrass representatives of each of the polynomials $\k_{ijk}x^ix^jx^k$. However, we again emphasize that two threefolds with the same $\calK$ are not necessarily Wall-equivalent. For instance, the cubics $8x_1^3+8x_2^3+x_3^3=0$ and $64x_1^3+x_2^3+x_3^3=0$ both have Weierstrass coefficients $(f,g) = (0,-27648)$, but two threefolds with these cubics cannot be Wall-equivalent, since the change of basis matrix that maps these cubics into each other, namely $\L=\operatorname{diag}(2,1/2,1)$,
has rational entries.

\subsubsection{Algebraic invariants}

In addition to the arithmetic point of view, the varieties
$\calK,\ \calI,\text{ and }\calH$
are objects in classical algebraic geometry. Consider any integer polynomial $f$, which in our applications will be $\k_{ijk}\, x^ix^jx^k$ or $H_{ijkl}\, x^ix^jx^kx^l$. From the symmetric $h^{1,1}\times h^{1,1}$ matrix of second derivatives,
\begin{subequations} \eq{h_{ij} = \partial_i \partial_j f\, ,} we can construct the Hessian \eq{h \coloneqq \det h_{ij}\, .}\label{eq:hessian}\end{subequations}
This is an integral polynomial, and so its \emph{integral content}, i.e.,~the gcd of all of its coefficients, is invariant under change of basis. Similarly, if $h$ factors, the \emph{shape} of its factorization, i.e.,~the number of factors of each degree, is an invariant. Thus, we have four Hessian invariants, two each from the Hessians of the defining polynomials of $\calK$ and $\calH$.

A richer set of invariants can be constructed from ideal-theoretic considerations. Begin with the ring \eq{R\coloneqq \bbZ[x_1,\ldots , x_{h^{1,1}}]} of polynomials in the $x_i$ with integer coefficients. An ideal $I$ of $R$ is a subgroup of the additive group of $R$ that is closed under multiplication by elements of $R$: for example, 
the subset of all polynomials such that each term includes at least one factor of $x_1$. For any ideal $I$ of $R$, we can define its saturation $I^{\sat}$ as the set of all polynomials that are in $I$ when multiplied by the $x_i$ to a sufficiently high power, i.e.,~we define
\eq{I^{\sat} \coloneqq \bigl\{f\in R \big| fx_i^N\in I\ \forall i \text{ for some N}\bigr\}\, .\label{eq:DefIsat}} $I^{\sat}$ is itself an ideal, and in particular contains $I$.%; as an ideal, it has a Gröbner basis, and we will use this Gröbner basis to compute invariants.

We will be largely concerned with ideals constructed from one of the varieties $\calK,\ \calI,$ and $\calH$  described above, so let us be more specific. Consider a variety $\calV$ defined by the simultaneous vanishing of a finite set $f_1,\cdots,f_n$ of polynomials; since in all of our examples the content of these polynomials is already included in the list of invariants, we are free to divide by the content, so without loss of generality we will assume that these polynomials $f_i$ have trivial content. We define the Jacobian ideal $J(\calV)$ of $\calV$ as the ideal generated by the $n$ polynomials and the order-$n$ minors of the $n\times h^{1,1}$ matrix $j_{ai} = \partial_i f_a$, i.e., \eq{J(\calV) \coloneqq \left<f_1,\cdots,f_n, \operatorname{minors}_n j_{ai}\right>.} We will usually drop the $\calV$, and refer to the Jacobian ideal just as $J$ for simplicity. We can construct its saturation $J^{\sat}$ as in equation~\eqref{eq:DefIsat}. Recall that elements of  $J^{\sat}$ are graded by their degree. Define the degree-$d$ content of the singular locus $M_d$ to be the gcd of the contents of all polynomials in $J^{\sat}$ of degree less than or equal to $d$. 

These invariants can be efficiently computed in terms of any generating set for $J^{\sat}$.  It is convenient to consider a Gr\"obner basis $G(J^{\sat})$ for $J^{\sat}$, which can be written schematically as \eq{G(J^{\sat}) = \left(g_m, g_{m-1}, \cdots, g_1, g_{0}\right),} where $g_{i}$ stands for the set of degree-$i$ polynomials in the Gr\"obner basis.

The degree-zero content of the singular locus, $M_0$, is simply $g_0$, and deserves special attention.\footnote{It is amusing to note that the integers $M_0$ can be quite large; the largest we have encountered in our dataset is 254754473628464286014940445745045420261589521392846806, which arises as the $M_0$ invariant of the $\calI$ associated to a phase at $h^{1,1}=5$.} The variety $\calV$ is smooth if and only if $M_0$ is nonzero.\footnote{For more details on singular loci and Gr\"obner bases, we refer to \cite{cox2013ideals}.}
Upon reducing $\calV$ over a finite field $\bbF_p$, $M_0$ itself gets reduced modulo $p$, and so the bad primes of a smooth variety $\calV$ are exactly those that divide $M_0$. As an integer that encodes the bad primes of a variety, $M_0$ is reminiscent of 
the conductor $N$ of an elliptic curve, as defined in e.g.~\cite{Ribet}.

Moving up in degree, the degree-one (or linear) content of the singular locus is given by \eq{M_1 = \gcd\bigl[\gcd\left(g_1\right),M_0]\bigr],} and it therefore divides $M_0$. The linear content is thus most useful in cases where $M_0$ vanishes, i.e., when $\calV$ is singular. Similarly, $M_2$ divides $M_1$, and so on. Eventually these invariants will  stabilize: there is some maximum $l$ such that $M_{k}=1$ for all $k\ge l$. In fact,  the $M_k$ can be defined in terms of any generating set of $J^{\sat}$, such as $J^{\sat}$ itself. Since $J^{\sat}$ contains $J$, and in particular contains the defining polynomials, which we have taken to all have trivial content, the degree-$d$ contents of the singular locus must stabilize
to one no later than the largest degree of the defining polynomial, but of course in any example they may stabilize well before this. Thus, to any variety $\calV$, and in particular to the varieties $\calK$, $\calI$, and $\calH$, we can associate the vector-valued content of the singular locus, 
$M$, defined as \eq{M := \Bigl(M_0, M_1, \cdots, M_{\operatorname{max}\left[\operatorname{deg}\left(f_a\right)\right]}\Bigr)\, .\label{eq:vector_mike_invariants}}

This completes the list of invariants  of the Wall data used in our analysis. We summarize these invariants in Table \ref{tab:wallinvts}.

\begin{table}[t!]
\begin{centering}
%\resizebox{\columnwidth}{!}{
\begin{tabular}{|c|c|c|c|}\hline
 & & &  \\[-1.0em]
Name & Label & Description & Equation\\[0.1em]\hline\hline
 & & &  \\[-1.0em]
Divisibility invariants & (a) & GCDs of combinations of $\k$ \& $c_2$& \eqref{eq:divinvd0}, \eqref{eq:hubsch} \\[0.1em]\hline
 & & &  \\[-1.0em]
Point counts & (b) & Number of points on varieties over $\mathbb{F}_{p^s}$ & \eqref{eq:ptcounts} \\[0.1em]\hline
 & & &  \\[-1.0em]
Hessians & (c) & Content \& shape of the Hessians & \eqref{eq:hessian} \\[0.1em]\hline
 & & &  \\[-1.0em]
Content of singular locus & (d) & Bad primes \& generalizations thereof & \eqref{eq:vector_mike_invariants} \\[0.1em]\hline
\end{tabular}
%}
\caption{A summary of the invariants of the Wall data used in the analysis. We will reference these labels throughout the text to describe which invariants are used in which stage of the analysis. } 
\label{tab:wallinvts}
\end{centering}
\end{table}

\subsubsection{Computing the invariants in an example}
 
Let us work through the computation of these invariants in an example. Consider the polytope whose vertices are given by the columns of the matrix $$\left( \begin{matrix} 1&0&0&1&-5\\0&1&1&0&-2\\0&0&2&1&-3\\0&0&0&2&-2\end{matrix}\right). $$
This polytope has a single FRST, which defines a toric variety whose anticanonical hypersurface is a threefold $X$ with $h^{1,1}=3$ and $h^{2,1}=43$.  We use the basis for $H_{2}(X,\bbZ)$ characterized by the rows of the GLSM charge matrix, given in this example by \eq{\left(\begin{matrix}
    -8&1&0&0&4&1&0&2\\0&0&1&1&0&0&0&-2\\-4&0&0&0&2&0&1&1
\end{matrix}\right)\, ,}
and use the dual basis for $H^2(X,\bbZ)$.

In this basis, the non-vanishing independent triple intersection numbers of $X$ are \eq{\k_{123}=1\, ,\quad  \k_{233}=-2\, ,\quad \k_{133}=2\, ,\quad \k_{333}=-8\, , \label{eq:exampleIntNums}}
%\sout{and cyclic permutations thereof,} and its 
and the second Chern class is given by \eq{c_2 = \left(12,12,4\right)^\top.} From these, one computes the divisibility invariants to be \eq{d_i = \left(4,  1,  2,  2, 16, 16, 16, 64\right).}
From the intersection numbers in Eq. \ref{eq:exampleIntNums}, we find that \eq{\k_{ijk}x^ix^jx^k = 6x_1x_2x_3 + 6x_1x_3^2 - 6x_2x_3^2 - 8x_3^3.} The Hessian is \eq{h = 432x_1x_2x_3 + 432x_1x_3^2 - 432x_2x_3^2,} so the Hessian content is 432. We consider the variety $\calK$ defined by \eq{3x_1x_2x_3 + 3x_1x_3^2-3x_2x_3^2 - 4x_3^3 \in \bbQ\bbP^2,} where we have scaled out the integral content of $\k_{ijk}x^ix^jx^k$. This variety is singular along the locus $x_2=x_3=0$, so we expect the integer content of the singular locus to vanish. Indeed, a generating set for $J^\sat$ is given by \eq{\left(x_3^3, 3x_1x_2, 3x_3\right),} so we find that the vector-valued content of the singular locus is given by \eq{M = \left(0, 3, 3, 1\right)\,.} Now let us compute the number of points on $\calK$ over $\bbF_2$; we readily check that the points in $\bbF_2^3$ satisfying the reduced defining polynomial \eq{x_1x_2x_3 + x_1x_3^2 + x_2x_3^2 =0} are\footnote{The origin is always excluded from point counts because the origin is not a point in projective space.} \eq{\left(x_1,x_2,x_3\right) = (0,0,1), (1,0,0), (0,1,0)\, , \quad\text{and}\quad  (1,1,0)\, ,} so we have \eq{\#\left(\calK,2\right) = 4\, .} Similarly, one computes \eq{\#\left(\calK,3\right) = 8 \quad \text{and} \quad \#\left(\calK,5\right) = 40\, .} 
The computation of invariants for $\calI$ and $\calH$ proceeds entirely analogously.

\subsection{Mori cone invariants}\label{sec:mori_invariants}

One further set of invariants does not manifestly involve the Wall data, but will be very useful in identifying (in)equivalent threefolds. Letting $X$ be a smooth --- but not necessarily simply connected or torsion-free --- threefold, we denote by $\mathcal{M}(X)\subset H_2(X,\mathbb{R})$ its Mori cone, generated by the effective curve classes in $H_2(X,\mathbb{Z})$. 

In general, the Mori cone of a threefold $X$ can depend on the choice of complex structure on $X$, i.e., it can enhance along certain loci of some non-zero codimension in the moduli space of complex structures of $X$. In order to avoid this subtlety, we assume $X$ has \emph{general} complex structure. Under this assumption, the Mori cone $\mathcal{M}(X)$ is itself an invariant of $X$, i.e., it is entirely determined by the diffeomorphism class of $X$.

Given the Mori cone $\mathcal{M}(X)$, and denoting $[\mathcal{C}^{(\alpha)}]\in H_2(X,\mathbb{Z})$, $\alpha=1,\ldots,l$ its extremal primitive generators, one can compute the integer-valued genus zero GV invariants
\begin{equation}\label{eq:GV_sequence}
   N_{\alpha,k}:=n^0_{k\cdot [\mathcal{C}^{(\alpha)}]}\, ,\quad k\in \mathbb{N}\, .
\end{equation}
The set of \emph{GV sequences} $N_{\alpha,k}$ is a powerful piece of data that allows comparing the diffeomorphism classes of pairs of threefolds to each other. Namely, given the Mori cones $\mathcal{M},\mathcal{M}^\prime$, and sets of GV sequences $N,N^\prime$ associated to a pair of threefolds $X$ and $X^\prime$, the threefolds are topologically distinct if the sets of invariants $N$ and $N^\prime$ do not agree. 

If $N$ and $N^\prime$ do agree, one only needs to consider any subset of $h^{1,1}$ Mori cone generators of $X$ that forms a basis, and enumerate all possible maps of these onto Mori cone generators of $X^\prime$ that share matching GV sequences. The subset of such maps that have determinant $\pm 1$ gives a finite (and for the Hodge numbers considered in this paper, typically small) set of candidate basis transformations $\Lambda\in \text{GL}(h^{1,1},\mathbb{Z})$. If any of them satisfies \eqref{eq:LambdaDef} then $X\simeq X^\prime$. Otherwise the threefolds reside in distinct diffeomorphism classes.

Computing the Mori cone of a threefold is in general difficult. However, Mori cones can be determined, in principle, via a computation of GV invariants as in \cite{Gendler:2022ztv}. Namely, one uses the fact that extremal rays of the Mori cone of a threefold $X$ come with non-zero GV sequences $N_{\alpha,k}$, with the following exception. 

Let $[\mathcal{C}]$ be a generator of $\mathcal{M}(X)$ such that an effective divisor $[D]\in H^2(X,\mathbb{Z})$ degenerates to a genus one curve worth of $A_1$-singularities when taking the K\"ahler class of $X$ to lie on the facet of the K\"ahler cone that is dual to the cone over $[\mathcal{C}]$. In this case, M-theory compactified on $X$ develops a non-abelian $\mathfrak{su}(2)$ enhancement of its generic $U(1)^{h^{1,1}}$ gauge group, with a single massless hypermultiplet in the adjoint representation \cite{Aspinwall:1995xy,Katz:1996ht}. The GV sequence of $[\mathcal{C}]$ is identically zero because the contribution to the index of the charged BPS vector multiplet is canceled by the charged hypermultiplets.

However, giving a vacuum expectation value (vev) to the adjoint scalar in the $\mathfrak{su}(2)$ vector multiplet (the Coulomb branch) amounts to moving back into the interior of the K\"ahler cone, breaking the gauge algebra back to its generic Cartan subalgebra. Likewise, giving a vev to the adjoint hypermultiplet (the Higgs branch) breaks the gauge algebra to the Cartan subalgebra. As the hypermultiplet moduli space is the complex structure moduli space of $X$ (plus Wilson line moduli of the M-theory three-form) the non-abelian enhancement never occurs for general complex structure. Rather, along the special locus in complex moduli space where the non-abelian enhancement can occur, a long vector multiplet breaks into a BPS vector multiplet as well as a BPS hypermultiplet, that can then become massless along a suitable locus in vector multiplet moduli space.

We thus conclude that the Mori cone of a threefold with \emph{general} complex structure\footnote{In Appendix \ref{app:superconfusingexample} we present a pair of hypersurfaces $X$, $X^\prime$ that are Wall-equivalent, but that have distinct Mori cones.  We show that this seeming contradiction is resolved by the fact that only one of the pair is general in complex structure moduli.}
is equal to the cone generated by all non-trivial GV sequences.
The above conclusion, reached by considering M-theory compactification on $X$, also follows from results of  P.M.H. Wilson (see the discussion at the beginning of Section~5 in \cite{wilson-GW} and also \cite{wilson-92, wilson-erratum, wilson-97}).
Specifically, Wilson showed that a (smooth projective) threefold $X$ of general complex structure does not contain a quasi-ruled surface $E$ over a smooth curve of genus $g > 0$ (that is, a surface $E \subset X$ that has a map $E \to C$, which is a conic bundle, with $C$ a smooth algebraic curve of genus $g \ge 1$). It follows that each extremal ray in the Mori cone of a threefold of general complex structure always contains a curve with nonzero GV invariant.\footnote{Likewise, 
for  general complex structure
the Weyl flops of genus $g>1$ described in \cite{Gendler:2022ztv} become ordinary flops, and the hyperextended K\"ahler cone becomes simply the extended K\"ahler cone.} 
 
Thus, by computing GV invariants to sufficiently high degree, e.g.~via the implementation of \cite{Demirtas:2023als} in \texttt{CYTools} \cite{Demirtas:2022hqf}, following \cite{Hosono:1993qy,Hosono:1994ax}, one can compute Mori cones. However, in practice one computes GV invariants to some specified cutoff degree, and some generators of the Mori cone might lie at higher degree. Thus, one only finds an \emph{inner} approximation of the Mori cone. 
Nevertheless, given a pair of threefolds, one can compute their GV invariants to some chosen cutoff degrees and run the algorithm described above under the assumption that the respective inner approximations of the Mori cones are exact. If in this way one finds a basis transformation  that satisfies \eqref{eq:LambdaDef}, one has proven that the threefolds lie in the same diffeomorphism class. Otherwise, the comparison is inconclusive.

In many cases, however, we can compute the exact Mori cone as follows: given a threefold $X$ realized as a hypersurface in a toric ambient fourfold $V$ (see the discussion in \S\ref{sec:toric}), the inclusion
\begin{equation}
    X\hookrightarrow V\, ,
\end{equation}
induces a natural inclusion of curve classes $\mathcal{M}(V)\subset \mathcal{M}(X)$. Thus, the Mori cone of the ambient variety provides an \emph{outer} approximation of the Mori cone of the hypersurface. By considering embeddings of the same threefold into distinct toric fourfolds\footnote{One way of getting distinct $V_I$ that give the same hypersurface $X$ is to consider toric fans induced from distinct fine, regular and star triangulations of a reflexive polytope that agree along two-faces. Moreover, whenever we find equivalences between threefold hypersurfaces, say by comparing GV invariants, we may add new Mori cone information to the intersection formula \eqref{eq:Mcap}.} $V_I$, $I=1,\ldots,N_V$, one can find better outer approximations of the Mori cone via the intersection
\begin{equation}\label{eq:Mcap}
    \mathcal{M}_{\cap}(X):=\bigcap_{I=1}^{N_V} \mathcal{M}(V_I)\, .
\end{equation}
Crucially, if the generators of $\mathcal{M}_\cap (X)$ have non-zero GV sequences, then $\mathcal{M}(X)\equiv \mathcal{M}_\cap (X)$.

Given a threefold $X$ whose Mori cone can be determined in this manner, one can conclusively compare it with all other threefolds $\hat{X}$ for which the subset of generators of $\mathcal{M}_{\cap}(\hat{X})$ hosting non-trivial GV sequences, together with the second Chern class $c_2(T\hat{X})$, span all of $H^2(\hat{X},\mathbb{R})$. In this case, one can again simply go through all possible ways that map these curve classes to counterparts of $X$ that share the same GV data.

\section{Equivalence classes of toric phases}
\label{sec:toric}

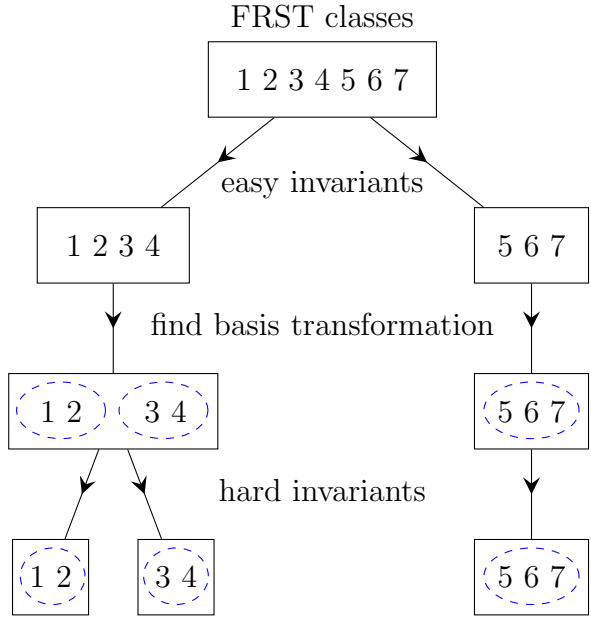
\begin{figure}[t!]
\begin{center}
\begin{tikzpicture}[scale=1.1]
\draw (0,6) node(1)[anchor=south,draw,black,minimum height=1cm,minimum width=3cm]{1 2 3 4 5 6 7}node[above=30pt]{FRST classes};
\draw (0,4.9) node(21)[anchor=south]{easy invariants};
\draw (-1.5-1,4.5-0.5) node(8)[anchor=south,draw,black,minimum height=1cm,minimum width=2cm]{1 2 3 4};
%\draw (-1.5-1-1.25,4.5-0.35) node(811)[anchor=south]{$\mathcal{S}_1$};
\draw (1.5+1,4.5-0.5) node(3)[anchor=south,draw,black,minimum height=1cm,minimum width=1.5cm]{5 6 7};
%\draw (1.5+1+1.1,4.5-0.35) node(311)[anchor=south]{$\mathcal{S}_2$};
\draw (0,5-1.75) node(22)[anchor=south]{find basis transformation};
\draw (-1.5-1,3-1) node(2)[anchor=south,draw,black,minimum height=1cm,minimum width=2.75cm]{1 2$\qquad$3 4};
%\draw (-1.5-1-1.6,3-0.85) node(811)[anchor=south]{$\mathcal{S}_1$};
\draw (-3.125,2.125) node(25)[anchor=south,draw,blue, ellipse,dashed,minimum height=0.75cm,minimum width = 1cm]{\hphantom{1 2}};
\draw (-1.9,2.125) node(26)[anchor=south,draw,blue, ellipse,dashed,minimum height=0.75cm,minimum width = 1cm]{\hphantom{3 4}};
\draw (1.5+1,3-1) node(7)[anchor=south,draw,black,minimum height=1cm,minimum width=1.5cm]{5 6 7};
%\draw (1.5+1+1.1,3-0.85) node(311)[anchor=south]{$\mathcal{S}_2$};
\draw (2.5,2.125) node(27)[anchor=south,draw,blue, ellipse,dashed,minimum height=0.75cm,minimum width = 1.25cm]{\hphantom{3 4}};
\draw (0,1.25) node(22)[anchor=south]{hard invariants};
\draw (-2.25-1,1.5-1.5) node(61)[anchor=south,draw,black,minimum height=1cm,minimum width=1cm]{1 2};
\draw (-0.75-1,1.5-1.5) node(62)[anchor=south,draw,black,minimum height=1cm,minimum width=1cm]{3 4};
\draw (-3.225,0.125) node(29)[anchor=south,draw,blue, ellipse,dashed,minimum height=0.75cm,minimum width = 0.85cm]{};
\draw (-1.75,0.125) node(30)[anchor=south,draw,blue, ellipse,dashed,minimum height=0.75cm,minimum width = 0.85cm]{};
\draw (1.5+1,1.5-1.5) node(5)[anchor=south,draw,black,minimum height=1cm,minimum width=1.5cm]{5 6 7};
%\draw (1.5+1+1.1,1.5-1.35) node(311)[anchor=south]{$\mathcal{S}_2$};
\draw (2.5,0.125) node(28)[anchor=south,draw,blue, ellipse,dashed,minimum height=0.75cm,minimum width = 1.25cm]{\hphantom{3 4}};
\draw[directed,black] (1) -- (3);
\draw[directed,black] (1) -- (8);
\draw[directed,black] (8) -- (2);
\draw[directed,black] (3) -- (7);
\draw[directed,black] (7) -- (5);
\draw[directed,black] (2) -- (61);
\draw[directed,black] (2) -- (62);
\end{tikzpicture}
\caption{A flowchart of our analysis for toric phases. Everything inside a rectangle is provably inequivalent to everything inside a different rectangle, and all phases within a blue oval are equivalent to each other.} 
\label{fig:flowchart}
\end{center}
\end{figure}

In this section, we describe our results for the classification of toric hypersurface threefolds, both simply connected and non-simply connected.  
The logic of our analysis is laid out in Figure \ref{fig:flowchart}. 

We construct Calabi-Yau threefolds as hypersurfaces in toric varieties following Batyrev's procedure \cite{Batyrev:1993oya}.\footnote{For more details, see the reviews in e.g.~\cite{Demirtas:2020dbm,Braun:2017nhi}.}
To this end, we pick a dual pair $(\Delta^\circ,\Delta)$ of four-dimensional reflexive polytopes from the Kreuzer-Skarke database \cite{Kreuzer:2000xy}. Then let $\cT$ be a fine, regular, star triangulation (FRST) of $\Delta^\circ$. The latter defines a fan for a toric variety in which a smooth threefold is obtained as the generic anticanonical hypersurface.

To count inequivalent toric threefolds, we first fix an $h^{1,1}$ and collect the full set of polytopes from the Kreuzer-Skarke list \cite{Kreuzer:2000xy} with this Hodge number. We then generate the list of triangulations of each polytope. For any given polytope $\Delta^{\circ}$, the full set of FRSTs is in general massively redundant: for FRSTs of four-dimensional reflexive polytopes, the triple intersection numbers and second Chern class depend only on the induced triangulations of two-faces. 
Hence, according to Wall's theorem, two FRSTs of $\Delta^{\circ}$ that have identical restrictions to two-faces
yield topologically equivalent threefold hypersurfaces (see~\cite{macfadden2023efficient} for recent work on generating such FRSTs).  Moreover, if $\Delta^{\circ}$ is invariant under any automorphisms, these produce a further redundancy: the associated action on the triangulations leads to trivial identifications of the corresponding threefolds.
We therefore work with \emph{FRST classes}, by which we mean sets of FRSTs that have identical restrictions to two-faces, up to the action of an automorphism of the polytope.
Counts of FRSTs and FRST classes are given in Table \ref{tab:toricSummary} for $h^{1,1}\leq 5$ and in Table \ref{tab:BoundsToricFav67} for $h^{1,1}=6,7$. From each FRST class, we construct a single threefold and compute its Wall data, i.e., its triple intersection numbers and second Chern class; these can be found efficiently with the \texttt{CYTools} software package \cite{Demirtas:2022hqf}. 

In this way we obtain a large list of threefolds at fixed $h^{1,1}$, and we aim to determine which of these threefolds are Wall-equivalent. Before attempting to prove equivalence between classes of Wall data, we first attempt to rule out the existence of such equivalences. As described above, we do this by computing invariants of the Wall data. We first sort the collection of threefolds into classes based on the invariants specified in Table~\ref{tab:wallinvts}. Elements of distinct classes are necessarily inequivalent, and so the remaining task amounts to a pairwise comparison among threefolds with the same sets of Wall invariants.

We begin by computing ``easy" invariants, such as $h^{2,1}$, the divisibility invariants, and point counts over finite fields of small prime order. Additionally, for toric hypersurface threefolds, we can sometimes provably construct their exact Mori cones, as explained in \S\ref{sec:mori_invariants}; in such cases, invariants such as the number of Mori generators and their GV invariants are also used. These invariants allow us to separate our list of threefolds into provably mutually inequivalent sets $\mathcal{S}$, each of which contains a number of potentially equivalent threefolds. 

Next, we attempt to find basis transformations between members of each set $\mathcal{S}$. We compile a list of candidate basis transformations by analyzing the generators of the Mori cone, as explained in \S\ref{sec:mori_invariants}. For instance, all generators with the same GV invariant must be mapped into each other. In this way, we can usually, but not always, find change of basis matrices between pairs of equivalent threefolds. We reiterate that, although GV invariants are used to guess the basis transformations, if an appropriate transformation is found, then the equivalence is \textit{proven}.

At this stage, inside each set $\calS$ we have constructed classes $\calC$ of threefolds that are provably equivalent to each other; these classes are denoted by blue dashed lines in Figure \ref{fig:flowchart}. What remains to do is to separate or merge the classes inside each set, which we refer to as indeterminate classes.

Actually, for the completed classifications presented in this paper, i.e.,~for 
%simply connected 
toric threefolds with $h^{1,1}\le5$, at this stage all equivalences have been found, and what remains to do is just to prove that all indeterminate classes are in fact inequivalent. The simplest way to do so is to compute computationally-intensive invariants of the Wall data, such as the Hessians, point counts over prime-power fields, and the vector-valued content of the singular locus. 

Even once all of the invariants listed in Table \ref{tab:wallinvts} are computed, it is still sometimes possible that there are classes that cannot be separated.\footnote{In our analysis, this happened only for $h^{1,1}\ge5$.} This usually occurs when there exists a change of basis matrix in $\operatorname{GL}(n,\bbQ)$ rather than $\operatorname{GL}(n,\bbZ)$ that maps the triple intersection numbers and Chern classes into each other. In this case, the two classes must be analyzed by hand, in a manner explained in Appendix \ref{app:excluding_basis_trafos}.

\begin{table}[t!]
\begin{centering}
\begin{tabular}{|c|c|c|c|c|}\hline
 &  &  & &  \\[-1.0em]
$h^{1,1}$ & {\#} polys & {\#} FRSTs & {\#} FRST classes & {\#} CYs  \\[0.1em]\hline\hline
 &  &  & &   \\[-1.0em]
6 & 16{,}608 & 584,281 & 74{,}503 & $54{,}025 \le \# \le 54{,}141$ \\[0.1em]\hline
 &  &  & &   \\[-1.0em]
7 & 48{,}221 & 5,990,333 & 467{,}283 & $ 337{,}620 \le \# \le 467{,}283$ \\[0.1em]\hline
\end{tabular}
\caption{
A summary of bounds for favorable threefold hypersurfaces in toric varieties at $h^{1,1}=6,7$.
In this table, all the numbers shown --- of polytopes, FRSTs, FRST classes, and topological equivalence classes of threefolds (denoted `CYs') --- refer to counts from \emph{favorable} polytopes.
For FRST classes at $h^{1,1}=6$, the invariants used, in the notation of Table \ref{tab:wallinvts}, 
were (a), (b) up to $\mathbb{F}_{23}$, (c), and (d), as well as $M_\cap$ invariants.
For $h^{1,1}=7$, the lower bound was derived using invariants (a),
(b)
up to $\mathbb{F}_7$,
and the content of the Hessian for $\kappa x^3$.}
\label{tab:BoundsToricFav67}
\end{centering}
\end{table}

We performed the analysis described above for 
all favorable and non-favorable polytopes with 
$1 \le h^{1,1} \le 5$.
We found that there are
$9444$  
equivalence classes of Wall data of the resulting toric threefolds.
Among these classes,  $9434$ have $\pi_1=0$, and have vanishing torsion in homology, and so Wall's theorem applies directly.
We have therefore obtained the \emph{exact numbers of topological equivalence classes of simply connected threefold hypersurfaces with $1 \le h^{1,1} \le 5$}; see 
Table \ref{tab:toricSummary}.

We now turn to threefolds that have $\pi_1 \neq 0$.
We find that the number of equivalence classes of Wall data is
1, 2, 3, 3, and 1 at $h^{1,1}=$ 1, 2, 3, 4, and 5, respectively.
All of these phases arise from triangulations of the 16 polytopes in the Batyrev-Kreuzer subset of the Kreuzer-Skarke list \cite{Batyrev:2005jc}.
If we were equipped with a generalization of Wall's theorem  establishing that non-simply connected threefolds with equivalent Wall data are homeomorphic, then the above numbers would be the numbers of topologically inequivalent Calabi-Yau threefold hypersurfaces with $\pi_1 \neq 0$.  

We note that the prepotential in type IIA string theory compactified on a threefold $X$ is determined by the Wall data of $X$.  Thus, given any two Wall-equivalent threefolds $X$ and $X'$, whether simply connected or not, type IIA compactifications on $X$ and $X'$ yield the same two-derivative effective theory for the vector multiplets.
It therefore seems plausible to us, on physical grounds, that Wall's theorem can be generalized to threefolds with $\pi_1 \neq 0$. 

Furthermore, we  emphasize that in every case we have found of a pair
$(X,X')$ of Calabi-Yau threefold hypersurfaces 
with equivalent Wall data, the fundamental groups are also isomorphic, i.e.,~$\pi_1(X)\simeq \pi_1(X')$. Thus, these pairs are diffeomorphic if and only if a) their universal covers $(\widehat{X},\widehat{X}^\prime)$ are diffeomorphic, and b) the freely acting symmetry groups $(G,G')$ relating
\begin{equation}
    X=\widehat{X}/G\, ,\quad  X'=\widehat{X}^\prime/G'\, ,
\end{equation}
are mapped into each other under the isomorphism induced from $\widehat{X}\simeq \widehat{X}^\prime$. We have shown that the first condition a) is satisfied for all pairs of non-simply connected pairs of Calabi-Yau threefold hypersurfaces in the Kreuzer-Skarke list, while we leave checking b) for future work.

At $h^{1,1}=6,7$ there are no longer any threefolds with $\pi_1 \neq 0$, or with torsion in homology,  
so to count topological equivalence classes it is sufficient to count equivalence classes of Wall data.
However, our approach becomes more expensive. For $h^{1,1}=6$, we included only threefolds obtained from triangulations of favorable polytopes. We computed both upper and lower bounds on the number of threefolds, but did not obtain the exact number. Finally, at $h^{1,1}=7$ we again considered only threefolds obtained from triangulations of favorable polytopes, but we performed only a partial analysis in which inexpensive invariants were computed and no attempt was made to find basis transformations.
We therefore computed a lower bound on the number of threefolds, but did not obtain an upper bound more constraining than that given by the number of classes of FRSTs. Our final results for $h^{1,1}\leq 5$ are laid out in Table~\ref{tab:toricSummary}, and those for $h^{1,1}=6,7$ appear in Table~\ref{tab:BoundsToricFav67}.

\section{Equivalence classes of non-toric phases}\label{sec:non-toric}

In this section we consider a broader class of threefolds that we term \emph{non-toric phases}.  By this we mean Calabi-Yau threefolds that are related by flops to toric phases, i.e.,~to Calabi-Yau threefold hypersurfaces in toric varieties defined by FRSTs of polytopes in the Kreuzer-Skarke list, but are themselves \emph{not} necessarily Wall-equivalent to any such toric phase.

Below we obtain counts of non-toric phases in the birational equivalence classes $[X]_{\text{br}}$ with $X$ any of the hypersurfaces in the Kreuzer-Skarke database with $h^{1,1}=2$ or $3$.  Unlike the counts of toric phases in the rest of the paper, the counts of non-toric phases given here are provisional, in a specific sense that we will explain after introducing our method.

First, we recall that all  threefold hypersurfaces in toric fourfolds defined by FRSTs of a fixed reflexive polytope $\Delta^\circ$ are birationally equivalent. One may relate any pair of such threefolds via a series of flop transitions, inherited from a series of bistellar flips from one FRST to the next. Across any elementary such flop transition, a set of exceptional $\mathbb{P}^1$s in the same homology class $[\mathcal{C}]$ shrinks to a set of conifold points, which get blown up by introducing another set of exceptional $\mathbb{P}^1$s in the class $-[\mathcal{C}]$. For flop transitions inherited from bistellar flips, one may describe the vanishing curves as a complete intersection of a pair of prime toric divisors --- associated with an edge of the triangulation interior to a two-face --- with the  hypersurface, and we will refer to such flops simply as \emph{toric flops}. As conifold singularities lie at finite distance in K\"ahler moduli space, all threefolds in the same birational equivalence class share a common moduli space of K\"ahler structures, called the \emph{extended K\"ahler cone} $\cK$.

We now combine the K\"ahler cones $\cK_{X}$ inherited from the toric fourfolds $V$ constructed from the set of all FRSTs $\mathscr{T}$ of a fixed polytope $\Delta^\circ$ and define
\begin{equation}
    \mathcal{K}_{\cup} := \bigcup_{\mathscr{T}(\Delta^{\circ})}~\cK_{X}\,.
\end{equation}
We thus obtain a sub-cone of the extended K\"ahler cone $\cK$, i.e.,~$$\cK_{\cup} \subseteq \cK\,.$$ 
%\nvg{possibly equal?}
The full extended K\"ahler cone $\cK$, however, may in general contain chambers given by K\"ahler cones of  threefolds that cannot be realized via a series of toric flops from any toric phase.  Rather, such threefolds are obtained via a series of flops, some of which are not toric. We will call such flops \emph{non-toric} flops, and, by slight abuse of language, call the resulting threefolds non-toric phases.

In \cite{Gendler:2022ztv}, an algorithm was presented to determine all members of a birational equivalence class $[X]_{\text{br}}$ by enumerating all flops (toric and non-toric), using the set of GV invariants of any representative $X$. We will utilize a version of this algorithm to construct,
for each threefold hypersurface $X$ with $h^{1,1}=2$ in the Kreuzer-Skarke database, a set of threefolds in $[X]_{\text{br}}$.  
Granting certain assumptions that we detail below, the sets we find are complete, i.e.,~in each case we construct all members of $[X]_{\text{br}}$.

To explain the required steps, we start by recalling the set of boundary phenomena associated with a facet of  the K\"ahler cone $\mathcal{K}_X$ of a threefold $X$. Assuming general complex structure, the following possibilities arise
\cite{wilson-92,Witten:1996qb}\footnote{Cases (a), (b), and (c) are referred to as limits of type I, III$_0$, and II, respectively, in \cite{wilson-92}.} (see also \cite{Gendler:2022ztv}):
\begin{enumerate}
    \item A flop transition leads into another chamber of the extended K\"ahler cone of $[X]_{\text{br}}$.  
    \item A holomorphic divisor $D$ degenerates into a rational curve.  
    \item A holomorphic divisor $D$ degenerates into an isolated point.
    \item The threefold degenerates into a lower-dimensional variety.
\end{enumerate}
Facets of the K\"ahler cone $\mathcal{K}_X$ of types (b),(c), and (d) 
are also facets of the extended K\"ahler cone, while case (a) only occurs on facets of $\mathcal{K}_X$ that are not facets of the extended K\"ahler cone. 

The first three limits arise at finite distance in moduli space, while the fourth one lies at infinite distance. Importantly, in all cases an effective curve $\mathcal{C}$ shrinks, and by inspecting the physical phenomena that arise upon compactification of M-theory on $X$ \cite{Gendler:2022ztv} one finds that the genus-zero GV invariant $n^0_{[\mathcal{C}]}$ is non-vanishing in all cases, as rigorously shown in \cite{wilson-GW}.\footnote{Note that the subtle case discussed in \cite{Gendler:2022ztv}, where a holomorphic divisor degenerates to a higher genus Riemann surface, does not arise for general complex structure \cite{wilson-GW}.}

As a consequence, by computing GV invariants\footnote{Efficient computation of GV invariants of hypersurfaces in toric varieties is possible in $\mathtt{CYTools}$, using the implementation \cite{Demirtas:2023als} of the classic construction \cite{Hosono:1993qy,Hosono:1994ax}.} of a threefold $X$ to sufficiently high degree, one can compute the Mori cone $\mathcal{M}_X$, as explained in \S\ref{sec:mori_invariants}.  By taking the dual of $\mathcal{M}_X$ one computes the K\"ahler cone $\mathcal{K}_X$. Denoting the primitive generators of the Mori cone by $[\mathcal{C}^{(\alpha)}]$, as in \S\ref{sec:mori_invariants}, one can determine which of the four phenomena listed above occur along the facet of $\mathcal{K}_X$ dual to $[\mathcal{C}^{(\alpha)}]$. 

First, one checks whether the facet
is of type (d): denoting by $\phi_\alpha$ the map
\begin{equation}
    H^2(X,\mathbb{R})\rightarrow \mathbb{R}\, , \quad [D]\mapsto \int_X [D]\wedge [\mathcal{C}^{(\alpha)}]\, ,
\end{equation}
an infinite distance limit (d) occurs whenever the triple intersection form $\kappa$ vanishes when restricted to the kernel of $\phi_{\alpha}$, because precisely in this case the Calabi-Yau volume vanishes along the corresponding facet of the K\"ahler cone. 

Otherwise, one proceeds to check if the facet is of type (c). For this we consider the Hessian of the cubic polynomial $f=\kappa_{abc}t^at^bt^c$. If
the Hessian
vanishes along the kernel of $\phi_\alpha$, then a divisor shrinks to a point, i.e.,~we have a limit of type (c) \cite{wilson-GW}.

The remaining two options, (a) and (b), are distinguished by the sequence of GV invariants $N_{\alpha,k}$ along multiples of $[\mathcal{C}^{(\alpha)}]$, cf.~equation~\eqref{eq:GV_sequence}.
For a flop transition (a), all GV invariants $N_{\alpha,k}$ are non-negative, and vanish for all $k>6$, but often already vanish for smaller $k$ \cite{katz1992gorenstein,Collinucci:2022rii}, 
while for a limit of type (b) we have
\begin{equation}
    N_{\alpha,k}=(2l,-2,0,\ldots) \quad \text{or} \quad N_{\alpha,k}=(-2,0,\ldots)\, ,
\end{equation}
for some integer $l\geq0$.

Given the above classification, we can construct the extended K\"ahler cone by successively continuing past K\"ahler cone facets of type (a), i.e.,~going through every possible flop.
Across a flop transition, the triple intersection form and the second Chern class transform as
\begin{align}
\kappa  \rightarrow \kappa -  n^0_{[\mathcal{C}]} \cdot [\mathcal{C}]\otimes [\mathcal{C}]\otimes [\mathcal{C}]\, , \quad 
c_2  \rightarrow c_2 + 2 n^0_{[\mathcal{C}]} \cdot [\mathcal{C}]\, ,
\end{align}
while the GV invariants also transform in a simple way \cite{Gendler:2022ztv}.

Therefore, given the GV invariants, computed to sufficiently high degree, of a  threefold hypersurface, one may successively identify flop transitions, and compute Mori cones, intersection numbers, and Chern classes of all chambers in the extended K\"ahler cone. 

As an aside, by systematically collecting all divisor classes that shrink along facets of type II, III$_0$ and along facets at infinite distance in moduli space, one  computes the effective cone (at general points in moduli space) of the birational equivalence class $[X]_{\text{br}}$ \cite{Alim:2021vhs}.

A potential subtlety in executing this algorithm arises when the number of chambers in the extended K\"ahler cone is infinite. Assuming finiteness of the number of diffeomorphism classes of  threefolds, as conjectured e.g.~in \cite{morrison1994kahler}, there must exist a fundamental domain inside $\mathcal{K}$ containing a finite number of chambers minimally representing the set of diffeomorphism classes contained in $[X]_{\text{br}}$. For the purposes of counting inequivalent threefolds, finding this fundamental domain is sufficient.

In practice, we will find the fundamental domain as follows: for any facet of a K\"ahler cone of type (a), we compute the dimension of the cokernel 
\begin{equation}
    \text{coker}\left(\kappa|_{\text{ker}(\phi_{\alpha})\times \text{ker}(\phi_{\alpha})}\right)\, ,
\end{equation}
where we view the triple intersection form as a map $\kappa:\, H^2(X,\mathbb{R})\times H^2(X,\mathbb{R})\rightarrow H_2(X,\mathbb{R})$. If the dimension of the cokernel is equal to one, there exists a unique, up to scale, divisor class $[D^{\alpha}]\in H^2(X,\mathbb{R})$ such that $\int_X J\wedge J\wedge [D^\alpha]=0$ for all K\"ahler classes on the facet of the K\"ahler cone dual to $[\mathcal{C}^{\alpha}]$. This allows defining a natural (Coxeter) reflection map $\Lambda: \,  H^2(X,\mathbb{R})\rightarrow H^2(X,\mathbb{R})$ defined via
\begin{equation}
    \Lambda^\alpha=\id-2\frac{[D^{\alpha}]\otimes [\mathcal{C}^\alpha]}{\langle [D^{\alpha}],[\mathcal{C}^\alpha]\rangle} \, ,
\end{equation}
that maps points in $\mathcal{K}_X$ to points beyond the flop facet. Often, in such a case, the Wall data of $X$ and its flopped phase turn out to be related by $\Lambda$, viewed as a linear change of basis transformation \cite{Lukas:2022crp}, and are thus in the same diffeomorphism class (provided they are simply connected and have torsion-free homology). A flop of this kind has been termed a \emph{symmetric flop} in \cite{Brodie:2021ain}. Passing through all flops that are \emph{not} symmetric flops, 
we always find
a finite number of chambers\footnote{All examples known to us of extended K\"ahler cones 
with infinitely many chambers
arise via infinite sequences of symmetric flops.
However, we are not aware of a proof that 
flops that are not symmetric flops cannot produce
infinitely many chambers.}
in $\mathcal{K}$, in which case these represent all diffeomorphism classes in $[X]_{\text{br}}$. 

Even following the above scheme, a crucial subtlety remains: it is in general hard to assess how many GV invariants have to be computed before the exact set of K\"ahler cones can be inferred. This subtlety is discussed in detail in \cite{Gendler:2022ztv}. Here, we take the following practical, though somewhat ad-hoc, approach:
\begin{itemize}
    \item We start with a threefold hypersurface $X$. We compute GV invariants using the \emph{degree method} of \cite{Demirtas:2023als}, where the cutoff degree is chosen to yield at least $N_{\text{min}}$ points in the Mori cone.
    \item We initiate the algorithm explained above.
    \item Along any extremal ray of a candidate Mori cone, we compute GV invariants up to a minimal multiple $d_{\text{min}}$ of the primitive generator $[\mathcal{C}]$, using the \emph{past light cone method} of \cite{Demirtas:2023als}. If the candidate ray was not an extremal ray of the true Mori cone to begin with, this frequently reveals the missing generators. If a new generator is revealed, we repeat this step, until the result is stable.
\end{itemize}
We perform these steps for all (favorable) 4d reflexive polytopes of fixed $h^{1,1}=1,2$ --- for each polytope $\Delta$ choosing $X$ to be the Calabi-Yau hypersurface in the toric fourfold defined via the Delaunay triangulation of $\Delta$. We then use our knowledge of the Mori cone to enumerate all diffeomorphism classes.

In order to assess whether or not enough GV invariants have been computed to correctly identify all Calabi-Yau threefolds, and their exact Mori cones, we compare the results of this classification (at fixed $h^{1,1}$) for increasing values of the ad-hoc parameters $(N_{\text{min}},d_{\text{min}})$. 
If our results remain stable beyond some threshold of these parameters, we deem the results robust.

Assuming the above algorithm yields the exact Mori cones of all threefolds in the birational classes of  threefold hypersurfaces of a given $h^{1,1}$, we can compute topological equivalence classes of threefolds, just as we did in \S\ref{sec:toric} for toric threefolds.
We first classify according to invariants defined in \S\ref{sec:wall_invts}, and finally compare the Mori cones of each pair in the same equivalence class of invariants computed as in \S\ref{sec:mori_invariants}.

\begin{table}[]
    \centering
    \begin{tabular}{|c | c | c | c | c|}
        \hline
        & & & & \\[-1.em]
        $h^{1,1}$  & \#(CY cl.) & \#(toric CY cl.) & \#(non-toric CY cl.) & \#(bir. equiv. cl.)\\[0.1em]\hline\hline
        & & & & \\[-1.em]
         $2$ & $35$ & $29$ & $6$ & $29$\\[0.1em]\hline
         %& & & & & \\[-1.em]
         %$3$ & $515$ &  &   &   & \\[0.1em]\hline
    \end{tabular}
    \caption{Results of computation of non-toric phases with $h^{1,1}=2$. We display the
    %total number of K\"ahler cone chambers identified, the 
    number of classes of Wall data identified, the number of such classes with a representative realized as a  threefold hypersurface in a toric fourfold, the number of classes without such a representative, and the total number of birational equivalence classes.}
    \label{tab:non-toric-results}
\end{table}

At $h^{1,1}=2$ we start with $N_{\text{min}}=100$ and $d_{min}=3$. We find that the resulting count of Calabi-Yau threefolds and their topological data is stable against increasing these parameters to $N_{\text{min}}=10,000$ and $d_{\text{min}}=10$.  The corresponding results appear in Table \ref{tab:non-toric-results}. 
These counts are provisional only in the sense that they rely on our computation of GV invariants having high enough degree so that the Mori cones we obtained are correct.

In contrast, at $h^{1,1}=3$, while 
we consistently find 
that the number of K\"ahler cone chambers is 515, %$N_{\text{chambers}}=515$, 
we have not reached a stable answer for the Mori cones of the outermost non-toric phases, and we therefore leave the full classification of these for future work.

\section{Conclusions}
\label{sec:conclusions}

In this work we have made progress in classifying and counting inequivalent Calabi-Yau threefolds constructed from the Kreuzer-Skarke list.  

Wall's theorem specifies topological data, which we called \emph{Wall data}, such that two simply connected threefolds $X$ and $X^\prime$ are diffeomorphic if and only if they have equivalent Wall data.  Computing the Wall data is straightforward, 
but to determine whether the Wall data for $X$ and $X^\prime$ are equivalent 
one generally has to perform a nontrivial integral change of basis relating $H_2(X,\mathbb{Z})$ to $H_2(X^\prime,\mathbb{Z})$.  Finding such a transformation, or showing that none exists, is a priori computationally intensive.

Our approach was to minimize 
the number of necessary comparisons by first classifying threefolds in terms of 
arithmetic and algebraic invariants of the Wall data.   
As one example, the triple intersection numbers determine a cubic surface $\mathcal{K} := \kappa_{ijk}x^i x^j x^k =0$, which is a variety defined over $\mathbb{Q}$, and 
the point counts on $\mathcal{K}$ over finite fields $\mathbb{F}_{p}$ and $\mathbb{F}_{p^s}$ (with $p$ a prime number, and $s$ a positive integer) are invariants of the Wall data.  
The full set of invariants we considered is listed in Table \ref{tab:wallinvts}.

Using these invariants, we identified equivalent and inequivalent threefolds.  That is, we were able to identify pairs of equivalent threefolds, exhibiting basis transformations through which their Wall data were demonstrably identical, and also inequivalent threefolds, for which we proved that no such transformation exists. 
Our analysis proceeded in several phases, depending on the type of threefold being considered. The different subanalyses, and the Wall invariants used in each, are summarized in Table \ref{tab:analyses}.\footnote{For favorable toric phases with $h^{1,1}\le4$, we obtained
the counts of inequivalent threefolds with  
two independent pipelines, one with and one without the $M_\cap$ invariants, with perfect agreement.}

\begin{table}[t!]
\centering
\resizebox{\columnwidth}{!}{
\begin{tabular}{|c|c|c|c|c|}
\hline
& & & & \\[-1.em]
Analysis & $h^{1,1}$ & Invariants used & Summary & Results \\[0.1em]\hline
& & & & \\[-1.em]
Toric favorable & $1\le h^{1,1}\le 5$ & (a)-(d) + $M_\cap$ & Table \ref{tab:toricSummary} & exact counts\\[0.1em]\hline
& & & & \\[-1.em]
Toric favorable & $h^{1,1}=6$ & (a)-(d) + $M_\cap$ & Table \ref{tab:BoundsToricFav67} & upper \& lower bounds \\[0.1em]\hline
& & & & \\[-1.em]
Toric favorable & $h^{1,1}=7$ & (a)-(c) & Table \ref{tab:BoundsToricFav67} & upper \& lower bounds \\[0.1em]\hline
& & & & \\[-1.em]
Toric non-favorable & $1\le h^{1,1}\le5$ & (a)-(d) & Table \ref{tab:toricSummary} & exact counts \\[0.1em]\hline
& & & & \\[-1.em]
Non-toric & $h^{1,1}=2$ & (a)-(d) & Table \ref{tab:non-toric-results} & provisional counts \\[0.1em]\hline
\end{tabular}
}
\caption{The various subanalyses presented in this paper. For each analysis, we describe the  invariants  used, following the notation in Table \ref{tab:wallinvts}, and give a reference to the appropriate summary table.}
\label{tab:analyses}
\end{table}
 
For simply connected Calabi-Yau threefold hypersurfaces with $h^{1,1} \le 5$, the above methods sufficed for us to give a complete classification and an exact count, as shown in Table \ref{tab:toricSummary}.
For $h^{1,1}$ = 6 and 7 we obtained bounds on the numbers of favorable threefolds: see Table \ref{tab:BoundsToricFav67}.

For non-simply connected threefolds, Wall's theorem as stated in \cite{wall} does not apply.
There are only 16 polytopes in the Kreuzer-Skarke list that yield non-simply connected threefolds, and for these cases we computed the Wall data, as well as the other invariants mentioned above.
The number of such classes with equivalent Wall data is 1, 2, 3, 3, and 1 at $h^{1,1}=$ 1, 2, 3, 4, and 5, respectively.

Finally, we extended our methods to non-toric phases. Using the algorithm laid out in \S\ref{sec:non-toric}, we constructed an ensemble of threefolds that are 
related to toric phases by flops that are not
bistellar flips. 
This ensemble is provisionally complete, in the sense that we repeated the computation of GV invariants at successively larger scales until the result was stable for multiple iterations. 
Within this set of threefolds, we identified topological equivalence classes using the same method used for the toric phases. The results of this classification can be found in Table \ref{tab:non-toric-results}.

We close with several questions for further work:
\begin{itemize}
\item Is there a single invariant that characterizes whether two phases are topologically equivalent? Here we used an ensemble of topological invariants, but one might hope that, as is the case for e.g.~Riemann surfaces, that there exists a unique invariant sufficiently powerful that it suffices to reproduce our analysis.
\item Given the Wall data of a threefold, is there an invariant that characterizes whether or not this data can be realized in a toric hypersurface?
\item Here we have studied the arithmetic geometry of the rational varieties associated to threefolds by their Wall data. Does the arithmetic of the Wall data  have any relationship to the arithmetic of the underlying threefolds? For instance, recall that, at $h^{1,1}=3$, the cubic variety $\calK$ is an elliptic curve if it is smooth; of the 186 phases at $h^{1,1}=3$, 60 of the associated cubic varieties are smooth elliptic curves with complex multiplication. Complex multiplication has appeared in the context of certain highly symmetric string compactifications, in e.g. \cite{DeWolfe:2004ns}. Do the phases whose cubic varieties have complex multiplication enjoy any special properties? 
\item We found that for toric phases at $h^{1,1}=5$, the number of diffeomorphism classes is $\approx 32\%$ smaller than the number of FRST classes.  How does this proportion change at larger $h^{1,1}$?
\item How numerous are non-toric phases in comparison to toric phases at large $h^{1,1}$?
\item It was shown in \cite{Demirtas:2020dbm} that an upper bound for the number of FRST classes of polytopes in the Kreuzer-Skarke list is dominated by FRSTs of a single polytope, which has Hodge numbers $h^{1,1}=491,\ h^{2,1}=11$. How many of these FRST classes give inequivalent phases? 
Is the total number of inequivalent toric hypersurface phases also dominated by triangulations of this polytope? For progress in enumerating  FRST classes of this polytope and their associated toric phases, see \cite{macfaddenstepniczka}.
\end{itemize}

\newpage
\section*{Acknowledgements}
We thank A.~Chandra, A.~Constantin, C.~Fraser-Taliente, T.~Harvey, and A.~Lukas for discussions during String Pheno 2023.  We thank T.~H\"ubsch for helpful correspondence. The research of NG, NM, LM, JM, and AS was supported in part by NSF grant PHY–2014071. RN is supported by a Klarman Fellowship at Cornell University. The work of NG was also supported in part by a grant from the Simons Foundation (602883,CV), the DellaPietra Foundation, and by the NSF grant PHY-2013858.
The work of MS was supported by NSF DMS-2001367 and a Simons Fellowship.

\appendix
\addtocontents{toc}{\protect\setcounter{tocdepth}{1}}

\section{Excluding basis transformations}
\label{app:excluding_basis_trafos}

Let $X$ and $X^\prime$ be threefolds with the same Hodge numbers, denoting $h^{1,1}(X)$ by $n$, and denoting by $c_2$, $c_2^\prime$ their second Chern classes, which are linear forms.  Similarly, let $\kappa, \kappa^\prime$ be the cubic intersection forms for $X, X^\prime$.

We wish to either find, or show non-existence of, an $n \times n$ integer matrix $\Lambda$ with determinant 1 or -1, which maps $c_2$ to $c_2^\prime$ and $\kappa$ to $\kappa^\prime$ (by means of equation \eqref{eq:LambdaDef}). We can do this using Gr\"obner bases and Macaulay2 \cite{M2} by creating an ansatz (method of undetermined coefficients).  We consider an $n \times n$ matrix with indeterminate entries, and consider the set of linear and cubic equations on these $n^2$ variables, arising from the constraints in \eqref{eq:LambdaDef}. For $h^{1,1} = 3$, this works extremely well, but for $h^{1,1} = 4$ and $h^{1,1}=5$ the Gr\"obner basis computations take a very long time.
 
To handle $h^{1,1} = 4$ and $h^{1,1}=5$, we consider the determinants $h, h^\prime$
of the Hessian matrices of $\kappa$, $\kappa^\prime$.  Each irreducible primitive (i.e., integer content one) factor of $h$ must map, up to a sign, to the irreducible factors of $h^\prime$ of the same degree.  If there are linear forms, this gives extra linear equations for the entries of $\Lambda$,  further reducing the dimensionality of the ansatz. Similarly, if the cubic equation $\kappa_{ijk}\, x^i x^j x^k = 0$ has singular points away from the origin, then these must be mapped to each other. This gives a finite number of possible candidate maps between the respective sets of singular points. For each such candidate, one gets an even further constrained ansatz for $\Lambda$. If for all such candidates there is no integer solution $\Lambda$ satisfying all the constraints, then the two threefolds are inequivalent. This reasoning works for checking inequivalence for all favorable toric threefolds with $h^{1,1}=4$ and $h^{1,1}=5$.  In the latter case, in seven out of the $8016$ favorable classes we rely on factors of the Hessian or singular points to prove inequivalence.

Another problem we encounter at $h^{1,1}\geq 3$ is to determine whether favorable threefolds can be equivalent to non-favorable threefolds. 
Let us take $h^{1,1} = 5$ as an example. The invariants we use break up the 134 non-favorable threefolds into 91 different sets, together with 77 favorable threefolds that might be equivalent to some of these. 
Many such pairs have factors of the Hessian and/or singular points, and we find matrices that determine equivalence as explained above. However, there are a few cases where the Hessian does not factor and there are no singular points, and in this last extremity we resort to a brute force method. There are 252 possible integer vectors of length 5, with entries 0, 1, -1, that can be columns of an invertible integer $5 \times 5$ matrix $G$.  We iterate through these, setting two of the columns of $G$ to be two of the 252 possibilities, and using the resulting ansatz on the remaining 15 variables.  Each time, we get an ideal in 15 variables, and we decompose this into irreducible components using Macaulay2 \cite{M2}. 
In this way, we find a map showing equivalence, in at most a few seconds per case.  
In particular, we find that in many cases there are equivalences between favorable and non-favorable threefolds.

\section{A curious pair of diffeomorphic threefolds}
\label{app:superconfusingexample}

In this appendix we consider a pair of Calabi-Yau threefolds $X$ and $X'$ that arise as hypersurfaces in toric fourfolds $V$ and $V'$ respectively. 
We show that $X$ and $X'$ have distinct K\"ahler, Mori and effective cones, but are nonetheless Wall-equivalent and hence in the same diffeomorphism class!  
As we will explain, the resolution of this seeming contradiction is that $X'$ is general in complex structure moduli, but $X$ is not.

\subsection{A Calabi-Yau threefold hypersurface $X$}\label{app:b1}
First we define the toric variety $V$. Let  $\Delta\subset N_{\mathbb{R}}:=N\otimes \mathbb{R}$ --- with $N\simeq \mathbb{Z}^4$ ---  be the reflexive four dimensional lattice polytope with points
\begin{equation}
	\begin{pmatrix}
		0 & p_1 & \cdots & p_8
	\end{pmatrix}=\begin{pmatrix}
		 0&  1& -4&  0&  0& -2&  0& -2& -1\\
		 0&  0& -2&  0&  1&  0&  0& -1&  0\\
		 0&  0&  0&  0&  0& -1&  1&  0&  0\\
		 0&  0& -1&  1&  0&  0&  0&  0&  0
	\end{pmatrix}\, .
\end{equation}
Let $\Sigma$ be the normal fan of $\Delta$ (ignoring the point $p_8$ interior to a facet of $\Delta$). We set $V_s:=\mathbb{P}_{\Delta}$ and partially desingularize $V\rightarrow V_s$ via a fine, regular and star triangulation (FRST). It turns out that $\Delta$ has two inequivalent FRSTs, $\mathcal{T}_1$ and $\mathcal{T}_2$, but their induced triangulations of the two-faces of $\Delta$ are equivalent. We will denote the two resulting toric varieties as $V_{1}$ and $V_{2}$. We denote their respective K\"ahler cones by $\mathcal{K}_1$ and $\mathcal{K}_2$.

We let $X$ be a generic Calabi-Yau hypersurface of $V_1$. Importantly, by varying the coefficients of the general anti-canonical polynomial in $V_1$ (and $V_2$) one parameterizes only a codimension-one sub-manifold of the complex structure moduli space of the resulting Calabi-Yau threefolds, so these surfaces are not general. 

Along the intersection of K\"ahler cones $\overline{\mathcal{K}_1}\cap \overline{\mathcal{K}_2}$ the hypersurface $X$ remains smooth because the resulting singular curve in $V_{1,2}$ does not intersect $X$. Hence the K\"ahler cone of $X$ contains the union of the K\"ahler cones of the two ambient varieties:
\begin{equation}
	\mathcal{K}_X\supset \mathcal{K}_{\cup}:=\mathcal{K}_1\cup (\overline{\mathcal{K}_1}\cap \overline{\mathcal{K}_2})\cup\mathcal{K}_2\, .
\end{equation}
We define $\mathcal{M}_\cap$ as the cone dual of $\mathcal{K}_{\cup}$, and we have $\mathcal{M}_X\subset \mathcal{M}_\cap$.

Next, we define a basis $\{[\mathcal{C}^a]\}_{a=1}^3$ of $H_2(X,\mathbb{Z})\simeq H_2(V,\mathbb{Z})$ via a basis of linear relations among the points not interior to facets $(p_1,\ldots,p_7)$,
\begin{equation}
	Q=\begin{pmatrix}
		2&  0&  0&  0& 1&  1& 0\\
		2&  0&  0& 1&  0&  0&  1\\
		2&  1&  1&  1& 0 &  0&  -1
	\end{pmatrix}\, ,
\end{equation}
encoding the intersection pairing ${Q^a}_i:=\mathcal{C}^a\cap D_i$ with $D_i$, $i=1,\ldots,7$, the prime toric divisors. We denote by $\{[H^a]\}_{a=1}^3$ the dual basis of $H^2(X,\mathbb{Z})\simeq H^2(V,\mathbb{Z})$. In this basis, the only non-vanishing triple intersection number is
\begin{equation}\label{eq:intersection_form_X}
	 \kappa_{123}:=\int_X [H_1]\wedge [H_2]\wedge [H_3]=2\, ,
\end{equation}
and the second Chern class is given by the curve class
\begin{equation}
	\vec{c}=\begin{pmatrix}
		24\\
		24\\
		24
	\end{pmatrix}\, ,\quad c_a:=\int_X c_2(TX)\wedge [H_a]\, .
\end{equation}
The extremal generators of the cone $\mathcal{M}_\cap$ are
\begin{equation}
	[\hat{\mathcal{C}^1}]:=[\mathcal{C}^1]=\begin{pmatrix}
		1\\0\\0
	\end{pmatrix}\, ,\quad 	[\hat{\mathcal{C}^2}]:=[\mathcal{C}^2]=\begin{pmatrix}
	0 \\ 1 \\ 0
	\end{pmatrix}\, ,\quad 	[\hat{\mathcal{C}^3}]:=-[\mathcal{C}^2]+[\mathcal{C}^3]=\begin{pmatrix}
	0\\-1\\1
	\end{pmatrix}\, . 
\end{equation}
Expanding the K\"ahler form of $X$ as $J=\sum_{a=1}^3 t^a [H_a]$ the cone $\mathcal{K}_{\cup}$ is defined by the linear constraints
\begin{equation}
	t^1>0\, ,\quad t^2>0\, ,\quad t^3>t^2\, .
\end{equation}
The Calabi-Yau volume is computed as
\begin{equation}
	\text{Vol}(X)=\frac{1}{6}\kappa_{abc}t^at^bt^c=2t^1t^2t^3\, ,
\end{equation}
and therefore $\text{Vol}(X)\rightarrow 0$ as $t^1\rightarrow 0$ and also as $t^2\rightarrow 0$. As a consequence, these facets of $\mathcal{K}_\cup$ are also facets of the K\"ahler cone of $X$. This can also be seen from the fact that the dual generators of $\mathcal{M}_\cap$ are proportional to effective curve classes in $X$ that can be represented as complete intersection curves
\begin{equation}
	2[\hat{\mathcal{C}}^1]=[\hat{D}_3\cap \hat{D}_4\cap X]\, ,\quad 2[\hat{\mathcal{C}}^2]=[\hat{D}_2\cap \hat{D}_5\cap X]\, .
\end{equation}
Along the third facet, where $t^2=t^3$, the Calabi-Yau volume remains finite, but the prime toric divisor $D_7:=\hat{D}_7\cap X$ degenerates into a curve of genus one. Indeed, its volume is computed by
\begin{equation}
	\text{Vol}(D_7)=\frac{1}{2}{Q^a}_7 \kappa_{abc} t^b t^c=2t^1(t^3-t^2)=2t^1\cdot \text{Vol}(\hat{\mathcal{C}}^3)\, ,
\end{equation}
and thus vanishes linearly as $\text{Vol}(\hat{\mathcal{C}}^3) \rightarrow 0$. This is in accordance with \cite{wilson-GW,wilson-92,wilson-erratum,wilson-97} because the Calabi-Yau hypersurfaces constructed above are not general in their complex structure moduli. One thus expects that the class $[D_7]$ ceases to be effective under a general deformation of complex structure. We will confirm this explicitly.

The curve class $[\hat{\mathcal{C}}^3]$ is likewise proportional to the class of an effective complete intersection curve
\begin{equation}
	2[\hat{\mathcal{C}}^3]=[\hat{D}_5\cap\hat{D}_7\cap X]\, .
\end{equation}
The shrinking divisor $D_7$ arises from the point $p_6$ interior to a one-face of $\Delta$. Setting $x_7=0$ we may set $x_4=1$, as $x_4x_7$ is in the Stanley-Reisner ideal of the toric ambient variety. Thus we obtain a toric description of the divisor $D_7$ as a hypersurface of bi-degree $(4,0)$ in the toric variety with scaling relations
\begin{equation}
	\begin{pmatrix}
		x_2 & x_3 & x_1 & x_5 & x_6\\
		\hline 0 & 0 & 2 & 1 & 1\\
		1 & 1 & 0 & 0 & 0
	\end{pmatrix}\, ,
\end{equation}
which identifies $D_7$ as a quartic curve in weighted projective space $\mathbb{P}_{[2,1,1]}$ times a $\mathbb{P}^1$ with homogeneous coordinates $[x_2:x_3]$. The fiber is an elliptic curve, and hence $D_7=T^2\times \mathbb{P}^1$. The complete intersection $\hat{D}_5\cap \hat{D}_7\cap X$ is a set of two distinct points inside the $T^2$ factor, and hence the class $[\hat{\mathcal{C}}^3]$ is in the same class as the $\mathbb{P}^1$ factor in $D_7$. In particular, as claimed above, the divisor $D_7$ degenerates into a singular curve of genus one.

In summary, we learn that $\mathcal{K}_X\equiv \mathcal{K}_{\cup}$ and $\mathcal{M}_X\equiv \mathcal{M}_\cap$ are smooth and simplicial cones. Along two of the facets of $\mathcal{K}_X$ the Calabi-Yau volume degenerates, while on a third facet the prime toric divisor $D_7$ shrinks to a genus one curve. Compactifying M-theory on $X$, along the facet where $D_7$ shrinks, one finds a non-abelian $\mathfrak{su}(2)$ enhancement of the gauge algebra with the field content of $\mathcal{N}=4$ Yang-Mills. The $\mathbb{Z}_2$ Weyl group of the Yang-Mills theory is generated by
\begin{equation}\label{eq:Weyl-group-X}
	w=\id-2\frac{[\hat{\mathcal{C}}^3]\otimes [D_7]}{\langle \hat{\mathcal{C}}^3, D_7 \rangle}=\begin{pmatrix}
		1 & 0 & 0\\
		0 & 0 &1\\
		0 & 1 & 0
	\end{pmatrix}\, ,
\end{equation}
acting on $H_2(X)$. Notably, it maps $[\hat{\mathcal{C}}^3]$ to minus itself and its conjugate action on $H^2(X)$ maps $[D_7]$ to minus itself.

Finally, we reiterate that due to a theorem by Wilson \cite{wilson-92}, the existence of the divisor $D_7$ --- a (trivial) $\mathbb{P}^1$ fibration over a genus one curve ---  implies that our hypersurface $X$ is not general, i.e.,~its embedding into the toric variety $V$ forces the complex structure onto a special sub-locus. Indeed, we have $h^{2,1}(X)=115$, while the number of monomial sections of the anti-canonical line bundle minus the dimension of the algebraic torus action on $V$ and minus one for overall scaling is equal to $114$.

\subsection{A Calabi-Yau threefold hypersurface $X'$}

Next, we follow analogous steps to construct $X'\subset V'$.\footnote{Primed quantities in this section refer to the Calabi-Yau threefold $X'$ as opposed to the threefold $X$ of the previous section.} The polytope $\Delta' \subset N_{\mathbb{R}}$ has lattice points  
\begin{equation}
	\begin{pmatrix}
		0 & p'_1 & \cdots & p'_8
	\end{pmatrix}=
	\begin{pmatrix}
		 0&  1& -2& -2& -2&  0&  0&  0& -1\\
		 0&  0& -1&  0&  0&  0&  0&  1&  0\\
		 0&  0&  0& -1&  0&  0&  1&  0&  0\\
		 0&  0&  0&  0& -1&  1&  0&  0&  0
	\end{pmatrix}\, ,
\end{equation}
and $p'_8$ is interior to a facet of $\Delta'$. There are three inequivalent FRSTs of $\Delta'$ (excluding $p'_8$), and again all of them agree on two-faces. Thus again, we can construct an inner approximation $\mathcal{K}'_\cup$ of the K\"ahler cone $\mathcal{K}_{X^\prime}$ of the Calabi-Yau threefold $X^\prime$, with  
\begin{equation}
	\mathcal{K}^\prime_\cup \subset \mathcal{K}_{X^\prime}\,  \qquad \text{and} \qquad \mathcal{M}^\prime_\cap \supset \mathcal{M}_{X^\prime}\,.
\end{equation}

A suitable basis of $H_2(X^\prime,\mathbb{Z})$ is defined by the rows of
\begin{equation}
	Q'=\begin{pmatrix}
		 2&  0&  1&  0&  0&  1&  0\\
		 2&  1&  0&  0&  0&  0&  1\\
		 2&  0&  0&  1&  1&  0&  0
	\end{pmatrix}\, .
\end{equation}
In this basis we compute the intersection numbers and second Chern class, and find the same result as in our previous example in \S\ref{app:b1}:
\begin{equation}
	\kappa'_{123}:=\int_{X^\prime} [H'_1]\wedge [H'_2]\wedge [H'_3]=2\, ,
\end{equation}
and
\begin{equation}
	\vec{c'}=\begin{pmatrix}
		24\\
		24\\
		24
	\end{pmatrix}\, ,\quad c'_a:=\int_{X^\prime} c_2(TX^\prime)\wedge [H'_a]\, .
\end{equation}
As both $X$ and $X'$ are simply connected, and have torsion-free homology, Wall's theorem implies that there exists a (generally not holomorphic) diffeomorphism
\begin{equation}
    \varphi:\, X\rightarrow X'\, .
\end{equation}
However, the generators of $\mathcal{M}'_\cap$ are now given by  the basis elements, i.e.
\begin{equation}
	[\hat{\mathcal{C}}'^1]:=[\mathcal{C}'^1]=\begin{pmatrix}
		1\\0\\0
	\end{pmatrix}\, ,\quad 	[\hat{\mathcal{C}}'^2]:=[\mathcal{C}'^2]=\begin{pmatrix}
		0 \\ 1 \\ 0
	\end{pmatrix}\, ,\quad 	[\hat{\mathcal{C}}'^3]:=[\mathcal{C}'^3]=\begin{pmatrix}
		0\\0\\1
	\end{pmatrix}\, ,
\end{equation}
and we will see momentarily that all three classes are proportional to the classes of complete intersection curves. We have
\begin{equation}
	2[\hat{\mathcal{C}}'^1]=[\hat{D}'_2\cap \hat{D}'_4\cap X^\prime]\, ,\quad 2[\hat{\mathcal{C}}'^2]=[\hat{D}'_3\cap \hat{D}'_4\cap X^\prime]\, , \quad 2[\hat{\mathcal{C}}'^3]=[\hat{D}'_2\cap \hat{D}'_7\cap X^\prime]\, ,
\end{equation}
and the Calabi-Yau volume vanishes along all three facets of $\mathcal{K}'_\cup$. We also note that the polytope $\Delta'$ has a symmetry that induces a $\mathbb{Z}_2$ symmetry on $H_2(X^\prime)$ that takes the same form as the Weyl symmetry generator in the example in \S\ref{app:b1}.

The crucial difference between this hypersurface $X'$, and the hypersurface $X$ constructed in \S\ref{app:b1}, is that $X'$ is \emph{general in complex structure moduli}, in the sense that all of complex structure moduli space of $X'$ is swept out by varying the coefficients of the anti-canonical polynomial. Indeed, as required by Wilson's theorem \cite{wilson-92}, for general moduli the divisor class  $\varphi_*([D_7])\subset X^\prime$ is not effective. Similarly, the generator $[\hat{\mathcal{C}}^3]$ of the Mori cone $\mathcal{M}_X$ does not map to an effective curve class under $\varphi_*$ for general moduli.

Finally, we note that the isomorphism mapping even-dimensional (co)homology groups of our two examples $X$ and $X'$ into each other maps the K\"ahler cone of $X'$ into the union of the K\"ahler cone of $X$ with its image under the Weyl group \eqref{eq:Weyl-group-X}.

\bibliographystyle{utphys}
\bibliography{refs}

\end{document}